\documentclass{aa} 
\usepackage{graphicx}
\usepackage{multirow}
\usepackage{subcaption}

\usepackage[colorlinks = true,
            linkcolor = blue,
            urlcolor  = blue,
            citecolor = blue]{hyperref}
\usepackage{txfonts,textcomp}
\usepackage{natbib}
\bibpunct{(}{)}{;}{a}{}{,}  
\usepackage{booktabs} 

\mathchardef\period=\mathcode`.
\DeclareMathSymbol{.}{\mathord}{letters}{"3B}

\def\aap{{\em A}\&{\em A}}

\def\apjl{{\em ApJ}}
\def\apjs{{\em ApJS}}

\defcitealias{Fernandez2021}{F21}
\defcitealias{Fernandez2024}{F24}

\begin{document}

\titlerunning{A catalog of ringed galaxies in the TNG50 simulation}
\authorrunning{Fernandez et al. 2024}

\title{A catalog of ringed galaxies in the TNG50 simulation: Analysis of their properties and structure}


\author{Julia Fernandez\inst{1}\thanks{\email{fernandezmjulia@unsj-cuim.edu.ar }}, Emanuel Sillero\inst{2,4}, Sol Alonso\inst{1} \& Patricia Tissera \inst{2,3,4}}

\institute{Departamento de Geof\'{i}sica y Astronom\'{i}a, CONICET, Facultad de Ciencias Exactas, F\'{i}sicas y Naturales, Universidad Nacional de San Juan, Av. Ignacio de la Roza 590 (O), J5402DCS, Rivadavia, San Juan, Argentina.
\and Instituto de Astrofísica, Pontificia Universidad Católica de Chile, Av. Vicuña Mackenna 4860, Santiago, Chile.
\and Centro de Astro-ingeniería, Pontificia Universidad Católica de Chile, Av. Vicuña Mackenna 4860, Santiago, Chile.
\and Millennium Nucleus ERIS
}
         
\date{Received xxx; accepted xxx}

\abstract
{} 
{We explore galaxies with ringed structures in the TNG50 simulation to study their frequency and the influence of physical processes on their properties and morphologies.}
{The catalog of ringed galaxies was compiled through visual classification of synthetic images from the TNG50 simulation. Galaxies were selected based on specific criteria: a redshift range of $0\period01 < z < 0\period1$, stellar mass $M_\star >10^9 M_\odot$, stellar half-mass radius $r_{50} > 1$ kpc, and a specific star formation rate (sSFR), $\rm{log(sSFR/yr}^{-1}) > -13$. Our classification allowed for differentiation between inner rings, outer rings,  combinations of rings -- inner + outer (i+o) --, and partial rings (pseudo-rings), including barred and non-barred ringed galaxies.
We constructed a suitable control sample of non-ringed galaxies with similar redshift, stellar mass, and environmental density distributions to those of the ringed ones. The comparison between the galaxies in both samples enabled the analysis of their properties and differences. Finally, we analyzed the surface density of stellar mass ($\Sigma_\ast(r)$) to better understand the distribution and structure of the rings in relation to the properties of their host galaxies.}
{We identified 807 ringed galaxies within the TNG50 simulation. Approximately 59\% of these galaxies possess an inner ring, 22\% a partial ring, 12\% an outer ring, and 7\% have i+o rings. Furthermore, our statistical analysis reveals that a notable 64\% (507 galaxies) exhibit bars. In terms of physical properties, we find that ringed galaxies exhibit a lower efficiency for star formation activity, reduced gas fractions, redder colors, and higher metallicities with respect to non-ringed disk objects. Moreover, ringed galaxies display greater variability in metallicity for a given stellar mass compared to their non-ringed counterparts, indicating distinct evolutionary processes.
From the analysis of radial profiles, we find that galaxies with outer rings exhibit a $r_{50}$ similar to or slightly larger than their control group, while those with inner or partial rings tend to have smaller sizes. A deeper exploration of radial density profiles revealed a pronounced central mass deficit preceding the ring structures, with inner and outer rings located approximately at $r_{50}$ and $1\period5 \, r_{50}$, respectively. Galaxies with both i+o rings have inner rings that are more compact and massive. Additionally, galaxies with partial rings exhibit deeper mass profiles than their controls, particularly in central areas where pseudo-rings extend beyond $r_{50}$. These findings not only improve our understanding of galactic evolution in ringed galaxies but also underline the complex interplay between mass distribution and galactic morphology.} 
{}

\keywords{galaxies: ringed structures - galaxies: fundamental parameters - galaxies: statistics}

\maketitle 
%

\section{Introduction}


Ringed galaxies provide valuable insights into the processes of galactic evolution, as these rings are believed to be closely associated with the secular evolution of galaxies \citep{Kormedy1979}. Several formation theories for ringed galaxies have been proposed, one of which involves the formation of rings linked to resonant interactions between the galactic bar and the disk material. As the bars rotate, they can trap stars and gas at resonant locations, creating high-density regions that appear as rings. Bars, being the main drivers of secular evolution 
\citep{Athanassoula2010}, induce gas inflow toward the central regions of the galaxy, promoting star formation (SF) and the development of nuclear rings \citep{Knapen2013}. Additionally, the interaction between bars and the disk can lead to the formation of inner and outer rings at specific resonance points, such as the inner 4:1 resonance (I4R) and the outer Lindblad resonance (OLR) \citep{Buta2017}. Some authors (\cite{Schwarz1981, Schwarz1984a, Schwarz1984b, Rautiainen2000}, and many others) have developed various models explaining the formation of gaseous ring structures around the Lindblad resonances. These models illustrate how a bar-shaped perturbation in the mass distribution can generate a spiral structure in the gas, which subsequently reorganizes the material in the galactic disk \citep{Buta1986}.

An alternative theory was proposed to explain the formation and properties of rings and spirals, particularly in barred galaxies. This theory suggests that the building blocks of these structures are chaotic orbits guided by manifolds associated with the Lagrangian points $L1$ and $L2$, located along the major axis of the bar. These manifolds can be considered as tubes that confine the orbits, allowing them to form thin structures in the configuration space \citep{RomeroGomez2007, Athanassoula2010}. On the other hand, multiple mechanisms for ring formation in galaxies lacking prominent barred structures have also been explored. Although bars have traditionally been considered essential for the accumulation and confinement of gas in resonant rings, some studies suggest viable alternatives. For example, it has been proposed that spiral density waves can drive gas toward specific resonances \citep{Mark1974,Grouchy2010}, albeit with less efficiency in the absence of bars, or that gravitational interaction with a nearby companion can mimic the effects of a bar, generating non-axial perturbations that result in ring formation \citep{Combes1988,Byrd1993, Knapen2004}. Alternatively, mild oval distortions in galaxies classified as SA can also facilitate ring formation, as observed in the case of NGC 7217, which exhibits an extremely axisymmetric morphology and yet possesses well-defined rings \citep{Buta1995B,VerdesM1995}. Another scenario for the presence of rings in galaxies without bars is that these rings could have originated during a period when the galaxy had a bar, which subsequently dissipated either due to tidal interactions with a neighboring galaxy or as a result of gas accretion leading to an increased central mass concentration \citep{Grouchy2005}. However, these ideas pose several challenges in understanding the long-term survival of ring structures. Consequently, some numerical simulations, particularly those that incorporate the dynamics of gas, suggest that bars may be transient features in the evolution of a galaxy. Significant gas inflows toward the galactic center or mergers with satellites can destroy these bars. Nonetheless, they can also regenerate as the cooling of spiral disks and the destabilization caused by new gas accretion create conducive conditions for their reformation \citep{Sellwood1984, Combes1991, Buta1995B, Buta1996}. It is important to note that although most ring structures are likely associated with resonances, not all rings originate from internal mechanisms. Some ringed galaxies, particularly those lacking bars, may form through more catastrophic processes such as galaxy collisions, as seen in collisional galaxies, or through accretion events, as in the case of accretion ringed galaxies or galaxies with polar rings \citep{Buta1996, Schweizer1987,Comeron2014,Smirnov2024}. For example, collisional ring galaxies, such as the well-known Cartwheel Galaxy, result from head-on collisions that generate expanding waves of SF, whereas polar rings, such as those seen in NGC 4650A, form from the accretion of material at high angles to the disk of the host galaxy \citep{Buta2011}.

Our understanding of ringed structures has advanced significantly due to a series of observational and theoretical studies that have shed new light on these phenomena. \cite{Gusev2003} analyzed NGC 2336, a ringed barred galaxy, and noted a star formation rate (SFR) typical of late-type spirals, along with outer disk colors characteristic of an older stellar system with minor contributions from younger stars. Furthermore, \cite{Grouchy2010} compared star formation rates (SFRs) in 18 barred and 26 non-barred ringed galaxies, finding that the star-forming activities in their inner rings were similar and independent of ring shape or bar strength. Further investigations into the physical characteristics of ringed galaxies have been conducted by \cite{Silchenko2018}, who focused on SF in early-type galaxies with outer rings. \cite{Buta2003} and \cite{Grouchy2008} explored the possible relationships between spiral arm winding, bar presence, and ring formation, and their effects on SF within inner rings. More recently, \cite{Gil2023} analyzed the star formation history (SFH) of the outer ring of NGC 1291, basing the study on integrated spectra and proposing that a gas-rich merger event is responsible for the observed distribution of stellar ages. Additionally, a correlation between the stellar populations of the outer ring and the inner region of the galaxy was highlighted, suggesting an inside-out formation linked to the assembly of the bars.

The environmental factors and how they influence the prevalence of ring galaxies have also been explored. 
\cite{Madore1980} and \cite{Elmegreen1992} found pronounced environmental influences, with rings more common in less companion-rich systems and the distribution of outer and pseudo-rings varying with environmental density. On the contrary, \cite{Wilman2012} and \cite{Buta2019}, who also conducted studies on the morphological-environment relationship, found no significant correlation between the presence of inner and outer rings or bars and the local environmental conditions.
Recent studies, such as those conducted by \cite{Fernandez2021} (hereafter \citetalias{Fernandez2021}), have demonstrated that ringed galaxies display behaviors distinct from those of normal spiral galaxies. Ringed galaxies tend to have lower SF activity, redder colors, and higher metallicity values compared to normal spirals. Environmental studies have reinforced these findings; \cite{Fernandez2024} studied ringed galaxies residing in both poor and rich groups. They found that while the frequency of ring formation remains consistent across different environments, the environmental density influences the SFRs, colors, and stellar population ages of these galaxies. In particular, denser environments tend to accentuate these observed properties.

Numerical simulations have been instrumental in enhancing our understanding of ringed galaxies over the years. These simulations have enabled researchers to model the complex gravitational and hydrodynamic processes within galaxies, providing a comprehensive view of how secular evolution shapes galactic structures on cosmological timescales. One of the foundational works in this field was conducted by \cite{Schwarz1981}, who analyzed how a gaseous disk responds to a rotating stellar bar using a simulation with 2000 gas particles. They discovered that, in the absence of an inner Lindblad resonance, the gas formed a spiral shock that evolved into a ring aligned with the stellar bar. Furthermore, it was observed that the continuous replenishment of gas could sustain this structure indefinitely. The models obtained showed a remarkable resemblance to the features of many real galaxies, both in the distribution of hydrogen and the velocity field. This work was later expanded on by authors such as \cite{CombesGerin1985, Byrd1994}, and \cite{Rautiainen2004}. \cite{buta2008} found that the dynamics of bars, including their strength and pattern speed, significantly influence the stability and longevity of rings, with stronger bars associated with more prominent and stable ring structures. Furthermore, \cite{Bagley2008}, in a study on early-type spiral galaxies, revealed that outer rings and pseudo-rings could form without gaseous dissipation, solely through disturbances from a growing gravitational bar. Additionally, \cite{Treuthardt2008} and \cite{Treuthardt2009} estimated the bar pattern speeds in the ringed galaxies NGC 1433, NGC 2523, and NGC 4245 using dynamical simulations.

Currently, there are several catalogs of ringed galaxies observed in different wavelengths. Among them, the Southern Ringed Galaxy Catalogue (CSRG, \cite{Buta1995}) and the Northern Ringed Galaxy Catalogue GZ2 (GZ2-CNRG, \cite{Buta2017}) stand out. These optical catalogs are designed to compare morphological predictions of test particle models \citep{Schwarz1981, Schwarz1984a, Schwarz1984b,  Byrd1994}. Additionally, researchers such as \cite{Nair2010} and \citetalias{Fernandez2021} created catalogs with detailed visual classifications of galaxies from the Sloan Digital Sky Survey (SDSS). Other studies, such as those by \cite{Buta2015} and \cite{Comeron2014}, went beyond optical observations to explore the characteristics of ringed galaxies in the near-infrared wavelengths. 

While numerous studies have explored various aspects of ringed galaxies, there remains a significant gap in the detailed examination of their properties within simulations and their comparison to observations. By addressing this gap, we aim to enhance our understanding of the formation and evolution of ringed galaxies. The primary aim of this study is to create a homogeneous and comprehensive catalog of resonant ringed galaxies in the TNG50 simulation. Then, we also intend to analyze the properties of these simulated ringed galaxies and compare them with observational data.

This article is structured as follows: Section 2 describes the database, the selection process for ringed galaxies within the TNG50 simulation, the specific criteria used to identify galaxies with rings, as well as bars, and a comparison with observational findings. Section 3 presents the criteria used to construct reliable control samples, comparing ringed and non-ringed galaxies in terms of stellar mass, redshift, and environmental density. Section 4 examines the properties of ringed galaxies, focusing on their SF activity, colors, and metallicity in comparison to non-ringed galaxies. Section 5 focuses on the analysis of radial profiles and their relationship with ringed structures. Finally, Section 6 provides a summary of the main conclusions derived from our analysis.

\begin{table*}
\centering
\caption{Numbers, percentages, and standard errors for galaxies with different types of rings in the analyzed $z$ for TNG50 and Sloan.} 
\begin{tabular}{@{}c|c c|c c@{}}
\toprule
\multirow{2}{*}{Ring Type} & \multicolumn{2}{c|}{Illustris TNG50} & \multicolumn{2}{c}{SDSS DR14} \\
\cline{2-5}
 & Number & Percentage & Number & Percentage \\
\midrule
\textsc{inner ring}          & 475 & 58.85$\pm$1.73\% & 857 & 48.77\%$\pm$1.19\% \\
\textsc{outer ring}          &  95 & 11.77$\pm$1.13\% & 186 & 10.58\%$\pm$0.73\% \\
\textsc{inner + outer rings} &  58 &  7.18$\pm$0.90\% & 372 & 21.17\%$\pm$0.97\% \\
\textsc{partial ring}        & 179 & 22.18$\pm$1.46\% & 342 & 19.46\%$\pm$0.94\% \\
\midrule
TOTAL & 807 & 100\% & 1757 & 100\% \\
\bottomrule
\end{tabular}
{\small}
\label{tab:combined_table}
\end{table*}

\begin{table*}
\centering
\caption{Numbers and percentages for the different types of rings in Illustris TNG50, for the available $z$ values.}
\begin{tabular}{@{}lrrrrrrrrr@{}}
\toprule
Ring Type        & z=0.01       & z=0.02       & z=0.03       & z=0.05       & z= 0.06      & z= 0.07      & z= 0.08      & z=0.1        \\ 
\midrule
Inner Ring       & 23 (74.19\%) & 27 (69.23\%) & 50 (70.42\%) & 65 (62.50\%) & 76 (58.46\%) & 70 (52.63\%) & 83 (55.70\%) & 81 (54.00\%) \\
Outer Ring       &  4 (12.90\%) &  4 (10.25\%) &  8 (11.26\%) & 11 (10.57\%) & 14 (10.76\%) & 18 (13.53\%) & 17 (11.40\%) & 19 (12.66\%) \\
Inner+Outer Ring &  3  (9.67\%) &  4 (10.25\%) &  4  (5.63\%) &  6  (5.76\%) &  6  (4.61\%) & 11  (8.27\%) & 11  (7.38\%) & 13  (8.66\%) \\
Partial Ring     &  1  (3.22\%) &  4 (10.25\%) &  9 (12.67\%) & 22 (21.15\%) & 34 (26.15\%) & 34 (25.56\%) & 38 (25.50\%) & 37 (24.66\%) \\ \bottomrule
\end{tabular}
\label{tab:table2}
\end{table*}

\begin{table}
\centering
\caption{Numbers and percentages of barred and non-barred ringed galaxies for the analyzed $z$ in Illustris TNG50.}
\begin{tabular}{c c c c c}
\toprule
 z & $\rm N_{barred}$ & Percentage & $\rm N_{non-barred}$ & Percentage \\
\midrule
\textsc{0.01} & 21 &  2.65\% & 10 & 1.26\% \\
\textsc{0.02} & 30 &  3.79\% &  8 & 1.01\% \\
\textsc{0.03} & 47 &  5.94\% & 24 & 3.03\% \\
\textsc{0.05} & 69 &  8.73\% & 33 & 4.17\% \\
\textsc{0.06} & 86 & 10.88\% & 42 & 5.31\% \\
\textsc{0.07} & 81 & 10.25\% & 49 & 6.20\% \\
\textsc{0.08} & 81 & 10.25\% & 61 & 7.72\% \\
\textsc{0.10} & 92 & 11.64\% & 56 & 7.08\% \\
\midrule
\textsc{TOTAL}&507&64.18\%&283&35.82\%\\
\bottomrule
\end{tabular}
{\small}
\label{tab:table3}
\end{table}

\begin{table*}
\centering
\caption{Numbers and percentages of barred galaxies with different types of rings for each analyzed $z$ in TNG50.}
\begin{tabular}{@{}lrrrrrrrr@{}}
\toprule
Ring Type         & z=0.01    & z=0.02     & z=0.03     & z=0.05     & z=0.06     & z=0.07     & z=0.08     & z=0.1     \\ 
\midrule
Inner Ring        &16 (76\%)  & 22 (73\%)  & 35 (74\%)  & 48 (70\%)  & 53 (62\%)  & 46 (57\%)  & 48 (59\%)  & 52 (56\%) \\
Outer Ring        & 3 (14\%)  &  3 (10\%)  &  5 (11\%)  &  8 (12\%)  &  6  (7\%)  &  9 (11\%)  &  8 (10\%)  &  9 (10\%) \\
Inner+Outer Rings & 2 (10\%)  &  3 (10\%)  &  3  (6\%)  &  3  (4\%)  &  5  (6\%)  &  6  (7\%)  &  8 (10\%)  & 10 (11\%) \\
Partial Ring      & 0 ( 0\%)  &  2  (7\%)  &  4  (9\%)  & 10 (14\%)  & 22 (25\%)  & 20 (25\%)  & 17 (21\%)  & 21 (23\%) \\
\bottomrule
\end{tabular}
\label{tab:table4}
\end{table*}

\section{Data: Catalog of ringed galaxies}

\subsection{The Illustris TNG50 simulation}

The IllustrisTNG\footnote{\url{https://www.tng-project.org/}} simulations represent a series of gravomagnetohydrodynamical models of the universe \citep{Weinberger2017, Pillepich2018}, executed using the AREPO code \citep{Springel2010} to simulate the joint evolution of dark matter (DM), baryonic matter, and supermassive black holes from high redshifts ($z=127$) to the present epoch ($z=0$). This suite consists of three main runs, TNG50, TNG100 and TNG300, representing three physical sizes of cosmological boxes (cubic volumes of 50 Mpc, 100 Mpc and 300 Mpc across), each designed to investigate different astrophysical phenomena with varying resolutions and volume scales. These runs are accompanied by
FoF\footnote{Standard Friends-of-Friends (FoF) algorithm with linking length b=0.2
. The FoF algorithm is run on the DM particles, and the other types (gas, stars, black holes) are attached to the same groups as their nearest DM particle.} 
halo and 
Subfind\footnote{The Subfind algorithm identifies gravitationally bound substructures considering all particle types and assigns them to subhalos/galaxies accordingly.} 
subhalo (galaxy) catalogs, merger trees, lower-resolution counterparts, and DM-only counterparts. All available for the 100 output snapshots of each simulation \citep{Nelson2019TNGDataR}.

For the purposes of this study we make use of the latest simulation in the project, TNG50 \citep{Nelson2019TNG50, Pillepich2019}, which is the highest-resolution simulation. TNG50, in particular, has been designed to bridge the gap between large-volume cosmological simulations and high-resolution simulations of individual galaxies. 
The initial gas content is sampled by $2160^3$ cells with a baryon mass of $8 \times 10^4 M_\odot$. Part of it makes up the interstellar medium (ISM) of about 20000 resolved galaxies with stellar mass $M_\star \gtrsim 10^7 M_\odot$, where the average spatial resolution of the gas for SF is approximately 100 to 140 parsecs.
TNG50 provides a finely detailed representation of the universe, enabling comprehensive investigations into the global and small-scale properties of galaxies with the ability to resolve their internal structural details along with their chemo-dynamic evolution.

\subsection{Selection criteria}

In our study, the identification of ring galaxies within the TNG50 simulation was conducted by selecting galaxies within a specific range of stellar mass and redshift. This range was chosen to align with the observed characteristics of galaxies in the \citetalias{Fernandez2021} catalog, which classified images from the SDSS Data Release 14 (SDSS-DR14) within a redshift range of $0\period 01 \leq z \leq 0\period 1$,  magnitude $g< 16 \period 0$, axial ratio $b/a > 0 \period 5$, and a concentration index value\footnote{$C=r_{90}/r_{50}$ is the ratio of Petrosian 90\%- 50\% r-band light radii} $C < 2\period8$ (see \citetalias{Fernandez2021} for more details). 

Specifically, our focus was on TNG50 disk galaxies with comparable stellar mass ($M_\star$) and a redshift interval of $0 \period 01 \leq z \leq 0 \period 1$, enabling direct comparisons with observations and prior studies of ringed galaxies. We used a total of eight snapshots corresponding to the redshift values $z=0 \period 01,\ 0\period02,\ 0\period03,\ 0\period05,\ 0\period06,\ 0\period07,\ 0\period08$ and $0\period1$, which are all the snapshots stored by the TNG50 run in the redshift range mentioned and allow a detailed analysis throughout it. Additionally, further criteria were established based on the $M_\star$, specific star formation rate (sSFR = $\rm{SFR}/M_\star$) and the structural size of the galaxies. Only galaxies with a $M_\star \ge 10^9 M_\odot$, a $\rm{log(sSFR/yr}^{-1}) > -13$ and a stellar half mass radius $r_{50} > 1$ kpc were included in the catalog. This threshold for the stellar mass ensures that the selected galaxies are resolved with at least 10000 stellar particles so the ring structures can be well  defined, the sSFR restriction enables that the selected galaxies maintain star-forming activity similar to the \citetalias{Fernandez2021} ringed galaxies sample. The requirement for the $r_{50}$ aims to select galaxies with a sufficiently extended structure to facilitate the clear identification of rings. 

\begin{figure}[t]
\centering
  \begin{subfigure}[b]{0.48\linewidth}
    \includegraphics[trim={0 30 0 30}, width=\linewidth]{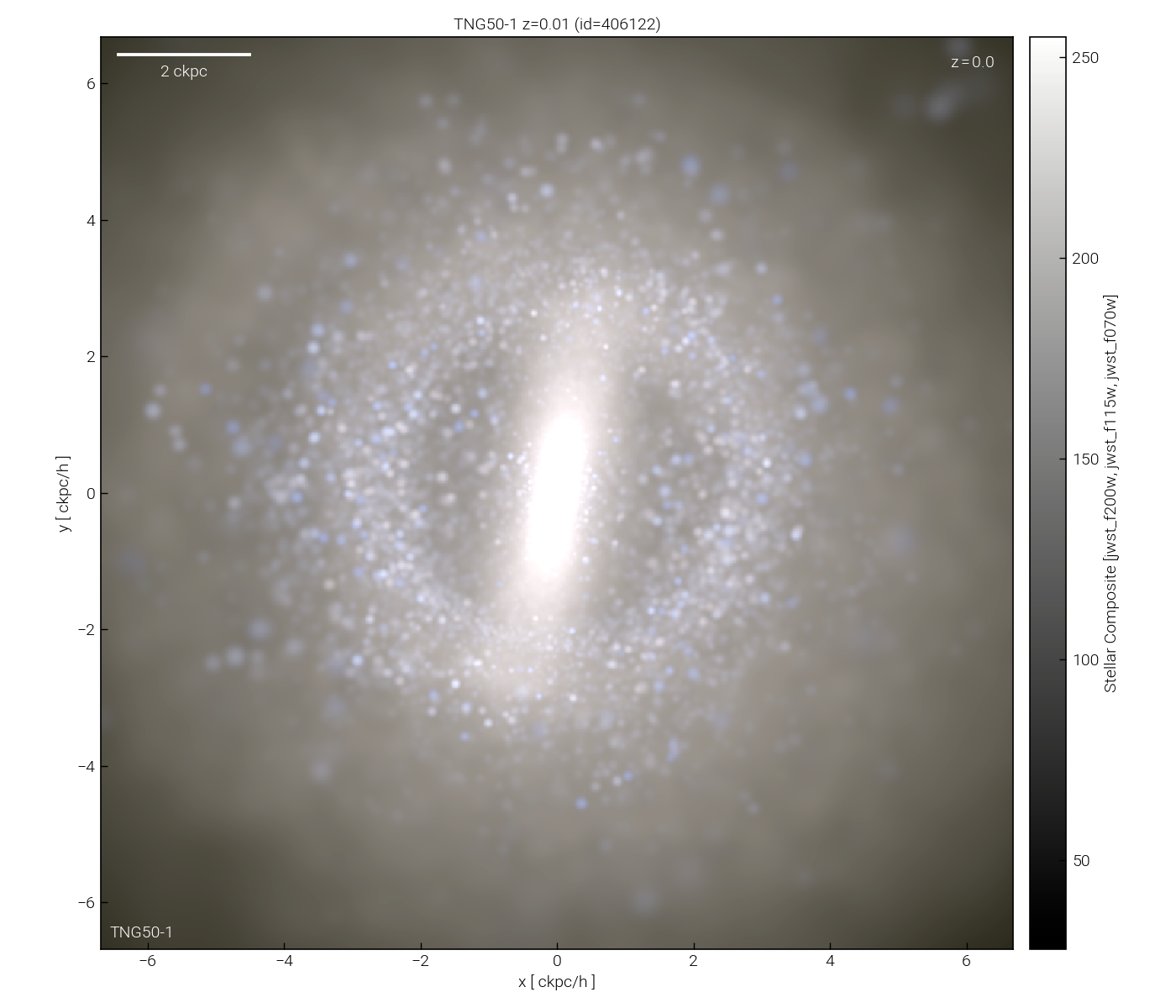}
    \caption{Inner Ring}
  \end{subfigure}
  \hfill
  \begin{subfigure}[b]{0.48\linewidth}
    \includegraphics[trim={0 30 0 30}, width=\linewidth]{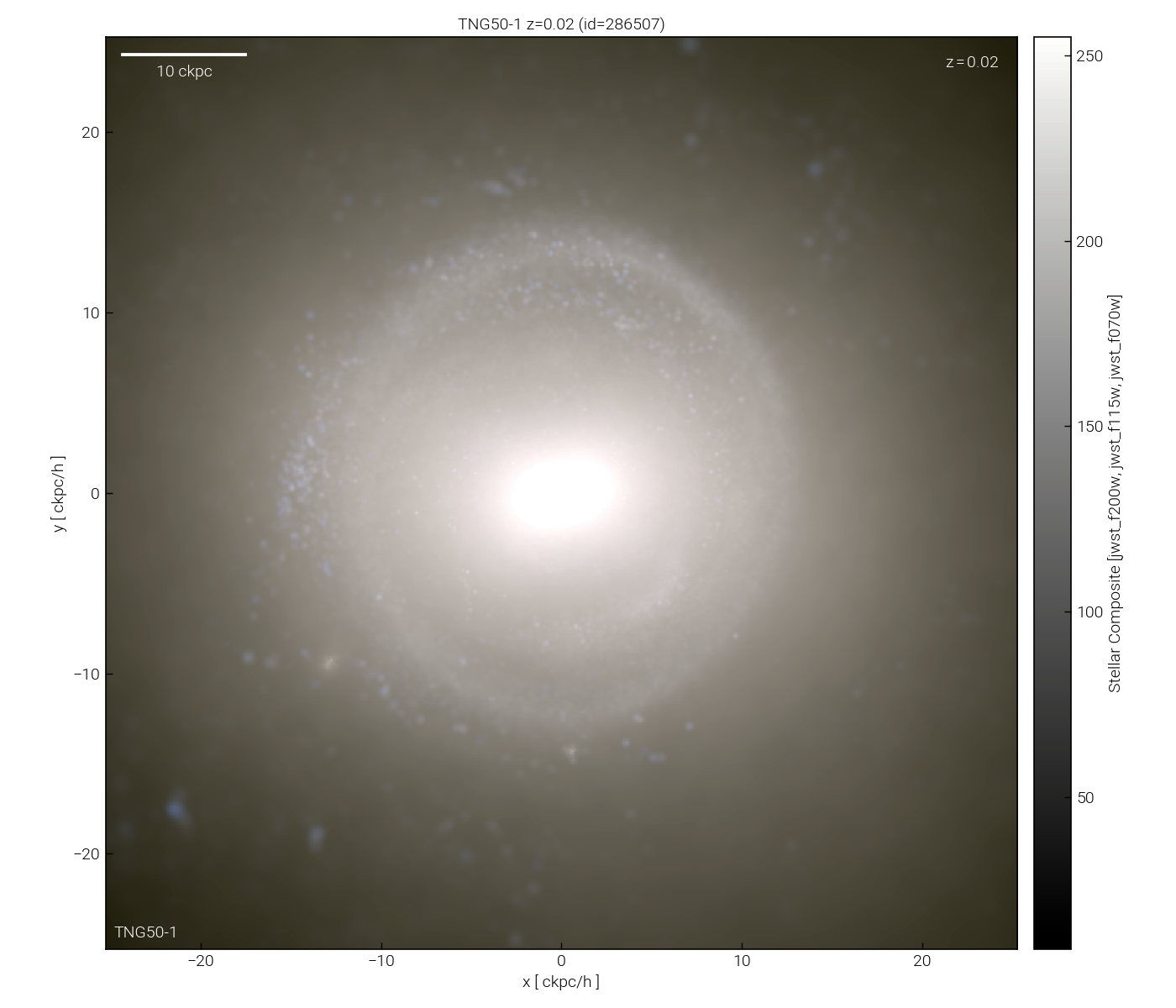}
    \caption{Outer Ring}
  \end{subfigure}
  
  \vspace{10pt} 
  
  \begin{subfigure}[b]{0.48\linewidth}
    \includegraphics[trim={0 30 0 30}, width=\linewidth]{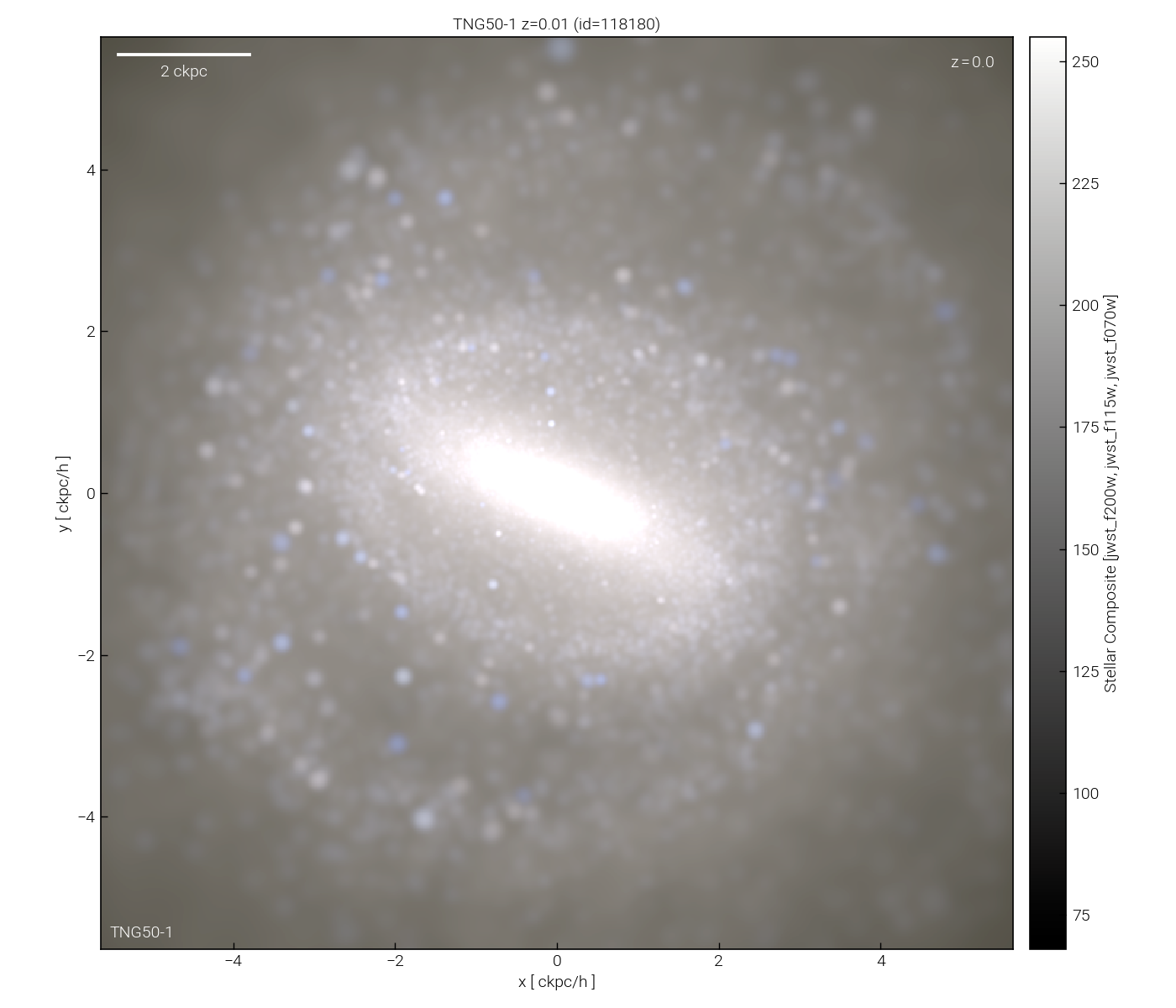}
    \caption{I+O Rings}
  \end{subfigure}
  \hfill
  \begin{subfigure}[b]{0.48\linewidth}
    \includegraphics[trim={0 30 0 30}, width=\linewidth]{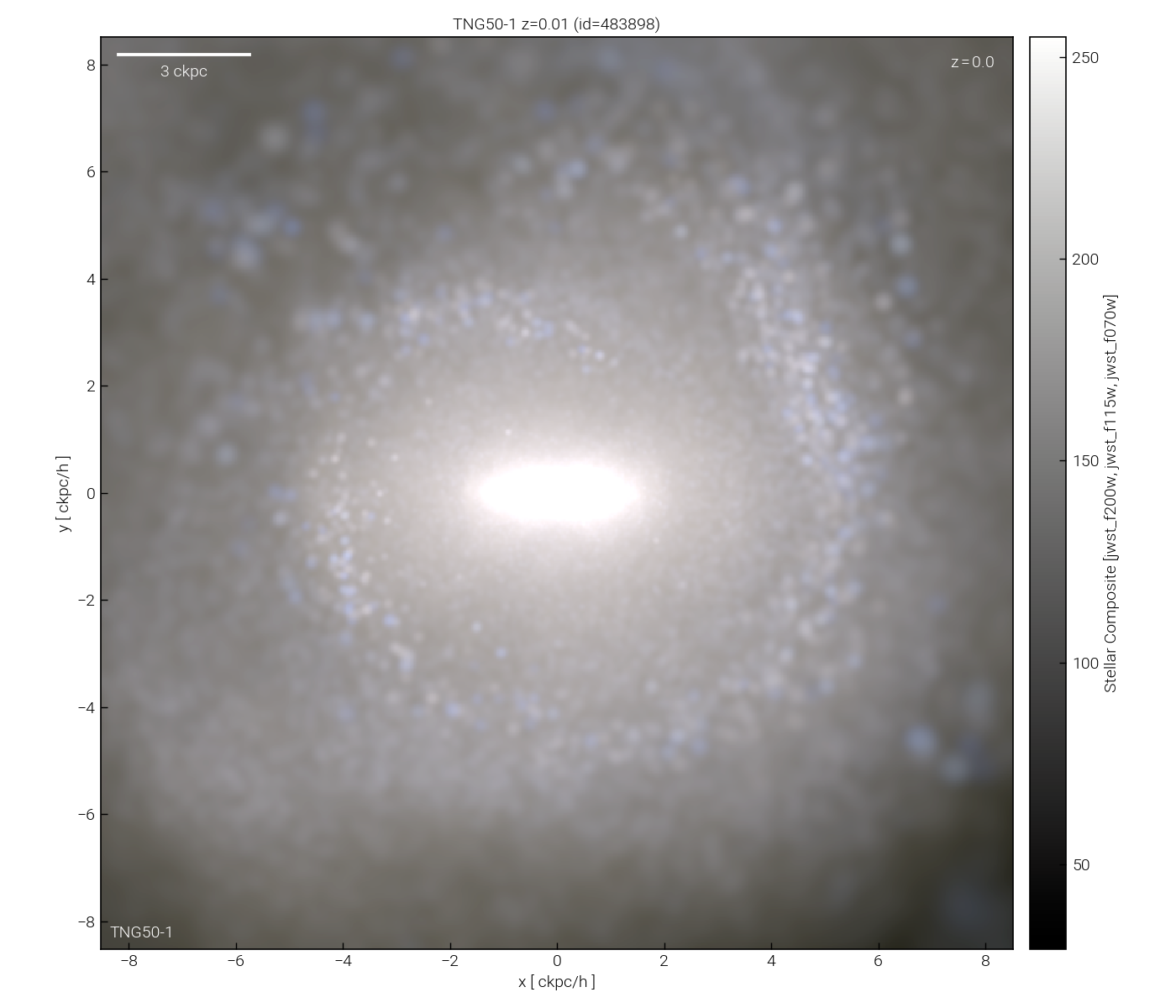}
    \caption{Partial Ring}
  \end{subfigure}
  
  \caption{Composite synthetic dust-free images of the different types of ringed galaxies found in TNG50, based on the simulated apparent F115W, F150W, and F200W JWST-NIRCam wide filter fluxes.}
  \label{fig:rings}
\end{figure}

\subsection{ Classification}

The visual classification of ringed galaxies within the TNG50 simulation was performed through the analysis of synthetic images, dust-free and in face-on projection, designed to simulate NIRCam and MIRI-like James Webb Space Telescope (JWST) observations of galaxies in apparent filter fluxes F070W, F090W, F115W. These images were constructed from light distributions of stellar populations and star-forming regions. The complete description of the synthetic images and measurements can be found in \cite{Vogelsberger2020}.

The principal parameter considered for our classification was the presence of ringed structures. The classification categories were based on the definitions established in previous studies by \citetalias{Fernandez2021} and \cite{Buta2017}, allowing a detailed differentiation between different types of rings.

The rings were classified into the following categories:

\begin{enumerate}
    \item Inner Ring: A closed circular or elliptical structure that envelops the bar or the central bulge of a galaxy.
    \item Outer Ring: A closed circular or elliptical external structure, typically twice the size of the bar (if present).
    \item Inner + Outer Rings (i+o): A combination of both inner and outer rings in the same galaxy.
    \item Partial or Pseudo-Ring: A circular or elliptical structure that is not completely closed, generally formed by spiral arms.
\end{enumerate}

Each type of ring was assigned a specific numerical value to facilitate its classification and statistical analysis. It should be noted that in the present work, galaxies with nuclear rings were not considered. Since they could bias our results as their classification might be overestimated due to the spatial resolution limit and/or the supermassive black hole (SMBH) feedback modeling in the simulation \citep[mainly for  galaxies with $M_\star \gtrsim 10^{10\period5} M_\odot$ -][]{Nelson2019TNG50}, among others. Therefore, we decided to investigate this type of ringed galaxies in a subsequent study. Figure \ref{fig:rings} displays representative images of the various types of ringed galaxies observed in the simulation, providing a clear visual reference that assists in illustrating the distinct structural characteristics defining each category. This facilitates a more intuitive understanding of the classification framework.

In accordance with the aforementioned process, we identified 807 ringed galaxies within the simulation that met the adopted criteria and form our target sample. Table \ref{tab:combined_table} displays the numbers, percentages and their associated standard error for ringed galaxies with different type of rings. Approximately, 59\% of these galaxies possess an inner ring, 22\% a partial ring, 12\% an outer ring, and 7\% i+o rings. Furthermore, Table \ref{tab:table2} shows the distribution of ring types across different redshifts. It is observed that at $z=0 \period 01$, galaxies with inner rings constitute the most significant fraction (74\%), whereas as redshift increases to $z=0 \period 1$, this fraction slightly declines to 54\%. Outer rings and i+o rings exhibit a relatively stable distribution across the available redshifts, implying a formation and evolution less time-dependent than inner rings. On the other hand, galaxies with partial rings appear to be more common as the redshift increases, starting with $3\%$ at $z=0 \period 01$ and growing to $25\%$ at $z=0\period 1$. This increase suggests that partial rings may not be permanent formations but rather transient features that emerge and disappear as a result of dynamic processes within galaxies over time. Such behavior would be consistent with the notion that these structures are more sensitive to interactions and the internal evolution of galaxies, thus reflecting a complex history of formation and destruction compared to more stable ring types. Additionally, their higher proportion at higher $z$ could suggest that these rings are more commonly found in the early evolutionary stages of galaxies, where the formation of more stable structures has not yet been completed. As galaxies evolve, partial rings may transform into more stable outer rings, reflecting a maturation process in the galactic structure, as identified in simulations conducted by \cite{Elmegreen1992}.

\subsection{Identifying bars in ringed galaxies}

The identification of bar structures in the ringed galaxies obtained from the TNG50 simulation was accomplished by correlating them with data provided by the Galaxy Morphologies (Kinematic) and Bar Properties catalog \citep{Zana2022}. This catalog contains kinematic decompositions of the stars of galaxies into different morphological components, as well as bar properties. Galaxies are decomposed into five morpho-kinematical components after an analysis of the stellar kinematics using the MORDOR code.
The data is available for TNG50 (all redshifts) and is restricted to subhalos with $M_\star \gtrsim 10^9 M_\odot$, corresponding to a minimum of $10^4$ star particles. 

Our research revealed that within the set of ringed galaxies identified in the TNG50 simulation, a notable 64\% (507 galaxies) exhibit bars, while the remaining 36\% (283 galaxies) do not. This suggests that the coexistence of bars in ringed galaxies is related to the formation or persistence of these rings in disk galaxies. Furthermore, as can be observed in Table \ref{tab:table3} there is a steady increase in the number and percentage of ringed galaxies with bars as z increases. This suggests that bar structures may be more consolidated or more easily identifiable at earlier stages of galaxy development in the simulation \citep{RosasGuevara}. In contrast, the percentage of galaxies with rings without bars does not show as pronounced an increase as that of galaxies with bars, reflecting more complex dynamics of bar and ring formation. 

Table \ref{tab:table4} details the presence of different types of rings in barred galaxies across the available redshift values. When comparing these data with Table \ref{tab:table2}, which represents the distribution of rings in all ringed galaxies regardless of the presence of bars, barred galaxies maintain a high proportion of inner rings at all analyzed z values, albeit with a less pronounced decrease compared to the general population. The fraction of these galaxies declines from 76\% to 57\% from $z=0\period01$ to $z=0\period1$. This could indicate that bars contribute to the stabilization of inner rings.
For outer rings and combinations of rings in barred galaxies, the proportions are comparable to those of the general population, suggesting that the influence of bars on the formation or maintenance of these rings may not be as critical. Furthermore, the proportion of partial rings in barred galaxies significantly increases with z, from 0\% to 23\%. This finding is consistent with the idea that, while bars may promote the formation of partial rings, these rings may be more transient or less stable over time.

\subsection{Comparison with the SDSS ringed galaxy catalog}

A comparison between our sample of ringed galaxies from the Illustris TNG50 simulation and the observational sample from the SDSS-DR14 (see \citetalias{Fernandez2021}) provides insight into the representativeness and accuracy of simulation models. Analyzing the data, we found notable differences as well as similarities in the distribution of ring types and the frequency of bars in both samples.

We found that for Illustris TNG50, inner rings are slightly more prevalent in the simulation at 59\% compared to 49\% in SDSS. However, the frequency of combined rings (i+o) is lower in TNG50 at 7\% than in SDSS at 21\%, which could suggest differences in underlying galactic dynamics or in the resolution of simulations and observations. Partial rings show a slight variation, being more common in TNG50 at 22\% compared to Sloan 19\%, while the outer rings present similar percentages in both samples (see Table 1).

Regarding the presence of bars in ringed galaxies, SDSS shows a slightly higher proportion with 68\% of ringed barred galaxies compared to 64\% in TNG50. This result reinforces the idea that bars are a prominent feature in ringed galaxies, both in the simulated and observed Universe. The slight discrepancy between the proportions could be attributed to differences in classification or even limitations in the simulation ability to accurately capture all the physical processes involved in bar formation. Despite the observed differences, the agreement in the general patterns of ring and bar classification suggests that TNG50 simulation captures fundamental aspects of galactic dynamics and evolution that define galactic morphology.

\section{Control sample}

\begin{figure*}[t]
    \center
    \includegraphics[trim={0 24 12 0},width=.335\textwidth]{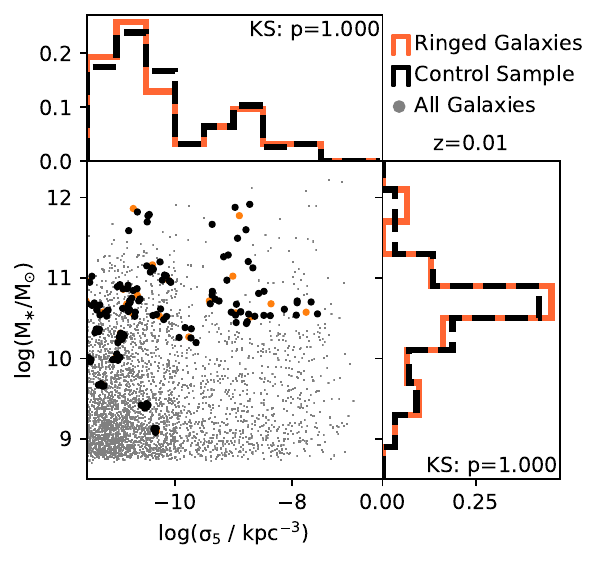}~\hfill 
    \includegraphics[trim={0 24 2 0},width=.333\textwidth]{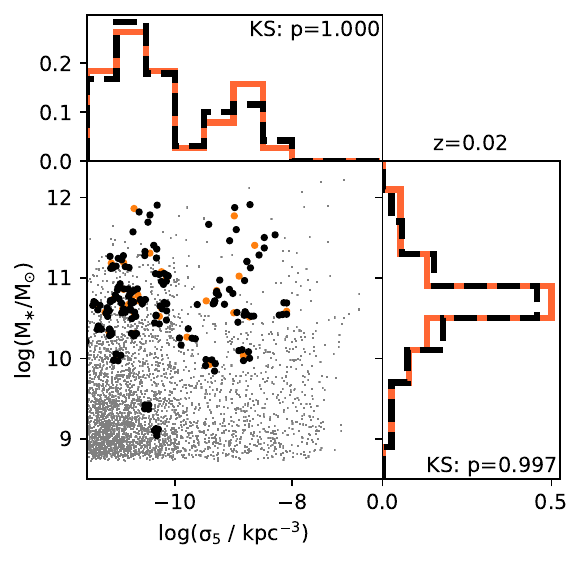}~\hfill 
    \includegraphics[trim={0 24 0 0},width=.333\textwidth]{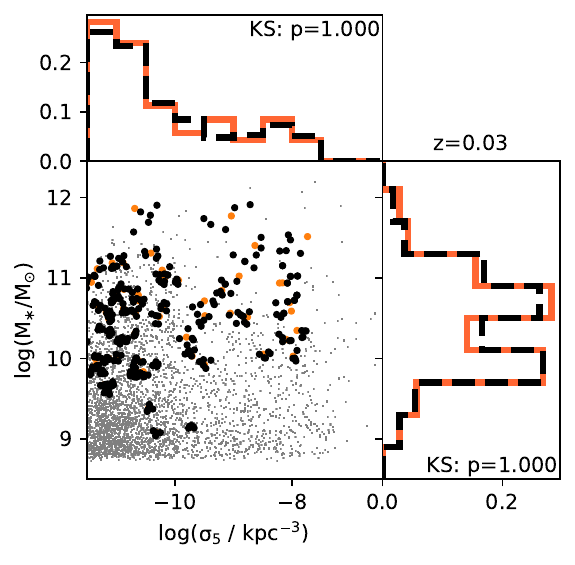}~\hfill \\
    \includegraphics[trim={0 24 0 0},width=.333\textwidth]{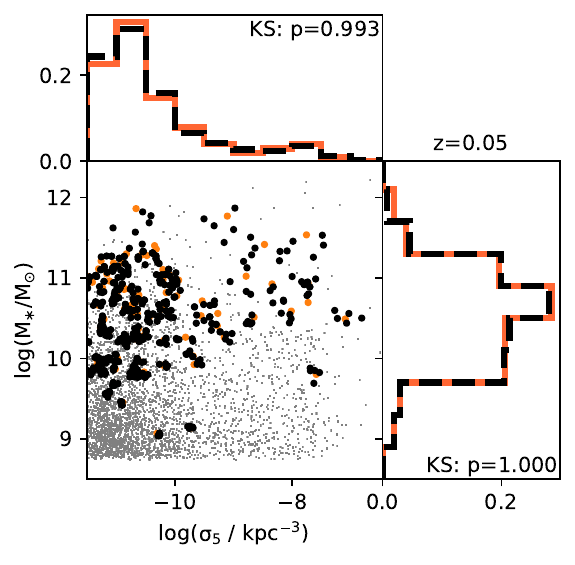}~\hfill 
    \includegraphics[trim={0 24 0 0},width=.333\textwidth]{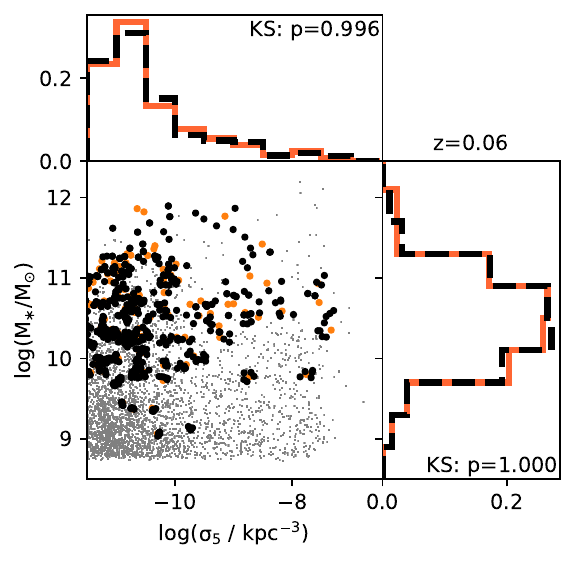}~\hfill
    \includegraphics[trim={0 24 0 0},width=.333\textwidth]{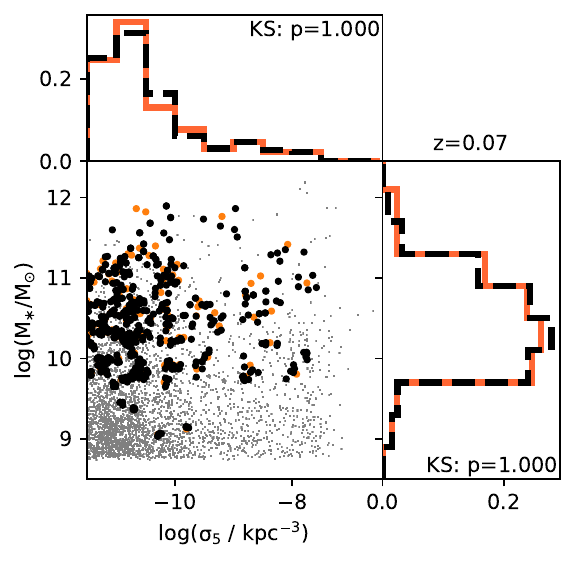}~\hfill \\ 
    \includegraphics[trim={0 10 0 0},width=.333\textwidth]{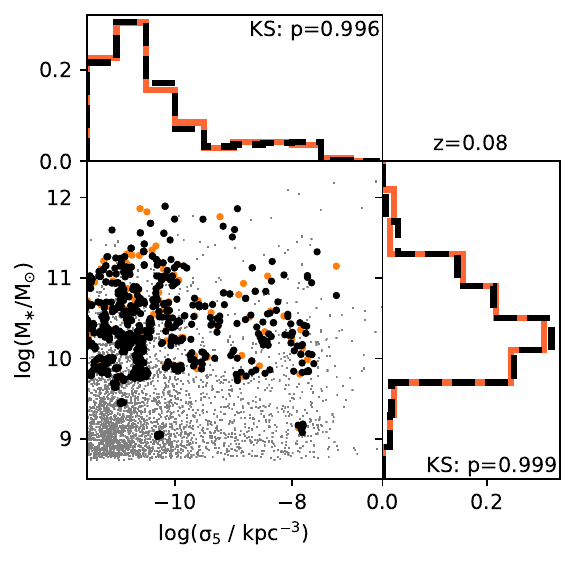}~\hfill 
    \includegraphics[trim={0 10 0 0},width=.333\textwidth]{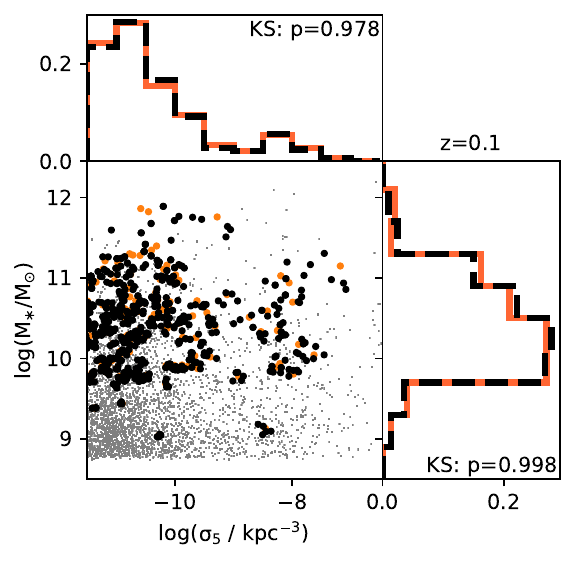}~\hfill
    \includegraphics[trim={0 10 0 0},width=.333\textwidth]{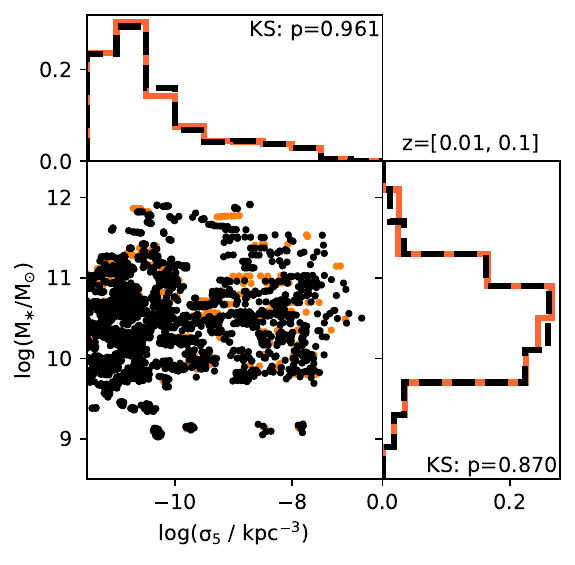}~
    \caption{Arrangement of galaxies with $M_\star > 10^9 M_\odot$ in the parameter space $\sigma_5 - M_\star$ (small gray dots). For each ringed galaxy (orange dot), its control group of non-ringed galaxies is shown (black dots, five nearest neighbors of equal z). The histograms show the distributions of each variable at each redshift for the ringed galaxies and their respective control samples. The $p$-values, obtained through a Kolmogorov-Smirnov (KS) test, representative of the goodness of fit of the distributions of both samples for each quantity, are displayed in their respective panels.}
    \label{fig:CS}
\end{figure*}

To accurately assess the relationship between rings and the properties of the host simulated galaxies, as well as to characterize these annular structures, we established a suitable control sample based on criteria analogous to those employed in \citetalias{Fernandez2021}.
We restricted our selection to simulated non-ringed disk galaxies that match the redshift, $M_\star$, and environmental density distributions \citep{Perez2009} of our target sample. We used $\sigma_{5} = 5 / (\frac{4\pi}{3} d_{5}^{3})$ as an indicator of the environmental density. This value was calculated for each galaxy in our sample, based on its distance to the fifth nearest galactic neighbor $(d_5)$, with $M_\star \ge 10^{9} M_\odot$.

To construct a control sample for each classified ringed galaxy, we selected its five closest non-ringed galaxies in the parameter space $\sigma_5 - M_\star$, located at the same redshift.
It is important to note that the neighbor search was conducted in a normalized parameter space, where each dimension was scaled to unity, to avoid numerical biases due to the magnitude scale of the different properties. A total of 4035 systems make up the general control sample.

Using the described method, we obtained control galaxy collections with z, $M_\star$ and $\sigma_5$ distributions similar to those of our ringed galaxy sample. This was corroborated by applying the Kolmogorov-Smirnov (KS) goodness-of-fit test, which yielded p-values $p>0 \period05$ for the null hypothesis at all redshifts, indicating that the target and control samples are subsets of the same distribution. Figure~\ref{fig:CS}, displays the ringed samples and their control ones for each analyzed $z$.

Thus, any observed differences in the features of the galaxies may be primarily associated with the presence of rings. Consequently, by comparing the results of the control sample with those of the ringed galaxies, we can explore how the processes responsible for ring formation correlate with the properties of the galaxies. Furthermore, this comparison allows us to disentangle the effects associated with rings from other influencing variables, thereby clarifying the underlying processes that affect factors such as SFR, color, metallicity, and overall galactic morphology.


\section{Ringed galaxy properties}

One of the primary objectives of our work is to compare the results obtained from simulations with observations by studying galaxies with different types of rings in terms of SF activity, color and metallicity, in comparison to the corresponding control sample. To achieve this, we divided our catalog of ringed galaxies into four specific simulated groups following \citetalias{Fernandez2021}: (i) ringed galaxies with an inner ring, (ii) ringed galaxies with other types of rings: double rings (i+o), outer rings, and partial rings, (iii) barred ringed galaxies, and (iv) unbarred ringed galaxies.

In the case of galaxies with inner rings and other types of rings, both groups may or may not exhibit bars. Similarly, barred or unbarred ringed galaxies may present any type of ring.

\subsection{Star formation activity}

 As is mentioned, in the TNG50 simulation, we quantitatively assess SF activity by considering for each galaxy its total stellar mass and its overall SFR. The former is obtained by summing the mass of all stellar particles assigned to the subhalo and the latter of all gas cells within the subhalo, respectively. In the present study, we will use the sSFR calculated from the aforementioned properties, providing a direct measure of the net rate at which new stars are forming from gas within these substructures.

\begin{figure}[t]
    \centering
    \includegraphics[width=.49\textwidth]{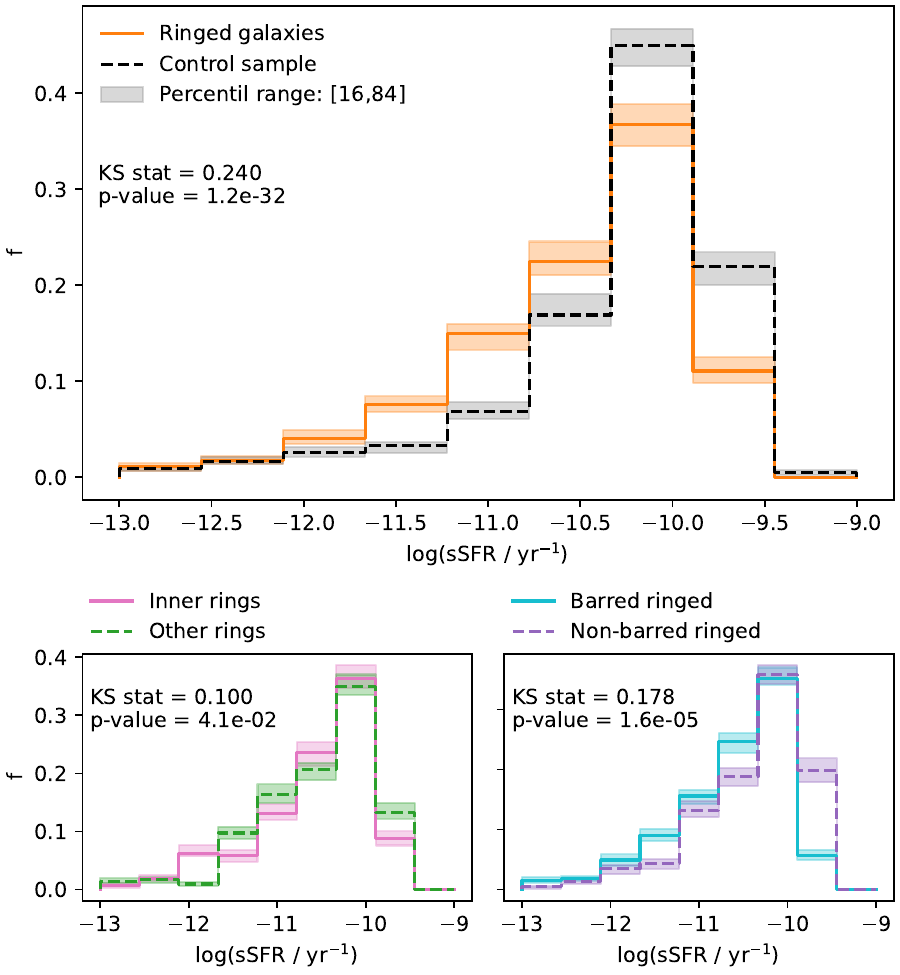}
    \caption{Distribution of the global log(sSFR) for the analyzed $z$ in the TNG50 simulation. The upper panel shows the fraction of ringed galaxies compared to the control sample. The lower panels provide a more detailed breakdown, showing the same relationship for inner ringed galaxies versus other types of rings (left panel) and barred versus non-barred ringed galaxies (right panel) across the available $z$ in TNG50.}
    \label{fig:sfr}
\end{figure}

Figure \ref{fig:sfr} presents the distributions of the global log(sSFR) for galaxies with rings and for the control sample. Additionally, bands representing the percentile range of [16, 84] has been included, capturing 68\% of the data dispersion. These are calculated using the bootstrap method and allowing the visualization of variability in the distributions due to sampling uncertainty. As observed in the figure, the distributions for both ringed galaxies and the control sample appear to be analogous. However, ringed galaxies exhibit a discernible shift toward lower SFR. This difference can be quantified using the KS-test, which reveals a significant discrepancy between the two distributions (KS-statistic=0.240, p-value=$1\period2\times10^{-32}$) and the lack of significant overlap between the bootstrap error bands further underscores the lower SFR in ringed galaxies compared to the control sample. In particular, the figure differentiates between galaxies with inner rings and other types of rings (lower left subpanel), as well as ringed galaxies with and without bars (lower right subpanel). In both cases, galaxies exhibit similar distributions. However, ringed galaxies with both an inner ring and a bar show a slight shift toward lower SFR compared to galaxies with other types of rings and those without bars. This observation is further corroborated by the KS-test results, which provide statistical support for the detected differences in SFR between these subgroups.

\begin{figure}[t]
    \centering
    \includegraphics[width=.495\textwidth]{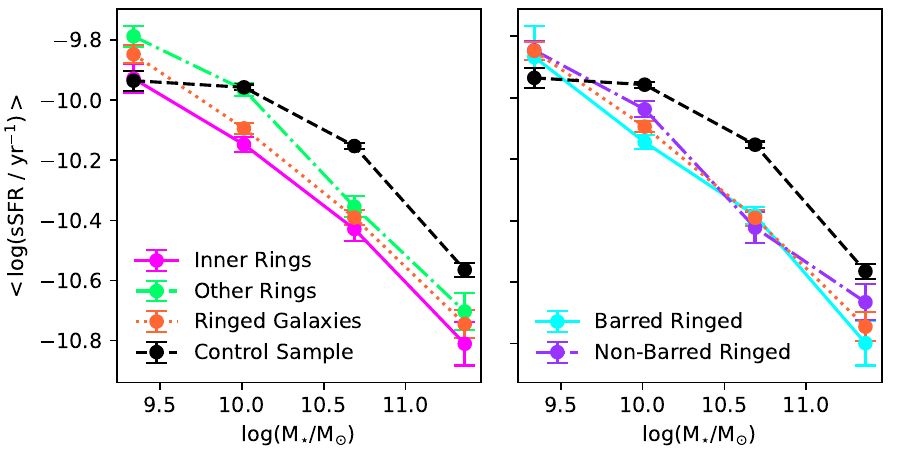}
    \caption{Mean log(sSFR) with associated bootstrap errors for five mass bins. Left panel: Mean values for all ringed galaxies, inner ringed galaxies, other types of ringed galaxies, and control sample galaxies. Right panel: Mean values for barred ringed galaxies and non-barred ringed galaxies.}
    \label{fig:sfr-masa}
\end{figure}

In Fig. \ref{fig:sfr-masa}, the analysis of the mean of the log(sSFR) and the associated bootstrap error for different mass bins provides additional insights into the influence of galaxies with structural features such as rings and bars on SF activity. As shown in the left panel of this figure it is evident that galaxies with inner rings tend to maintain lower sSFR compared to those with other types of rings and the general ringed population, indicating that the presence of inner rings might be associated with a depletion of the star-forming activity. While certain research indicates that SFR in outer rings are low \citep{Kostiuk2016, Katkov2022}, we observed that this property is even more pronounced in inner rings. Previous studies, such as those by \cite{Grouchy2010}, reported that while bars might concentrate gas in resonant regions and theoretically promote SF in inner rings, they did not necessarily increase the overall SFR. Furthermore, the right panel of Fig. \ref{fig:sfr-masa} displays the mean values and their associated bootstrap errors for barred ringed galaxies and non-barred ringed ones. It can be appreciated that barred ringed galaxies exhibit a systematically lower sSFR across almost the entire mass range compared to their non-barred counterparts. This result is similar to that found by \citetalias{Fernandez2021} and aligns with the findings by \cite{Ashley}, which illustrate how the suppression of stellar activity may be related to morphological and mass factors. This underscores the importance of considering a broader range of variables, such as the internal structure and mass composition of galaxies, when assessing the impact on the dynamics of SF. 

The gas fraction, $f_{\rm gas}$, within galaxies, is defined as $f_{\rm gas} = \frac{M_{\rm gas}}{M_{\rm gas} + M_\star}$, where $M_{\rm gas}$ is the mass of gas and $M_\star$ is the mass of stars. This parameter serves to quantify the evolutionary stage of a galaxy. 
In the context of the TNG50 simulation, analyzing $f_{\rm gas}$ across different galaxy types provides insights into how galaxies convert their gas into stars over time. A lower $f_{\rm gas}$ typically indicates a more evolved galaxy, where a significant fraction of the gas has been transformed into stellar content. Conversely, a high $f_{\rm gas}$ suggests a younger or less evolved galaxy, with abundant gas available to fuel future SF. 

\begin{figure}[t]
    \centering
    \includegraphics[width=.49\textwidth]{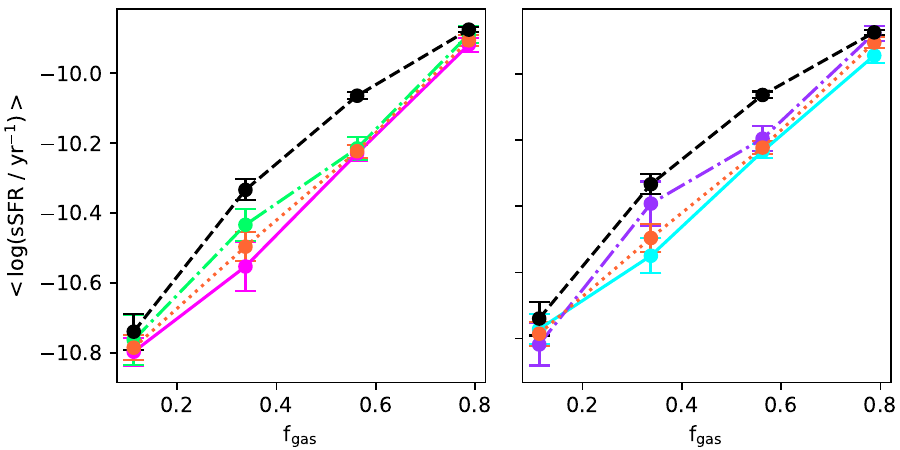}
    \includegraphics[width=.49\textwidth]{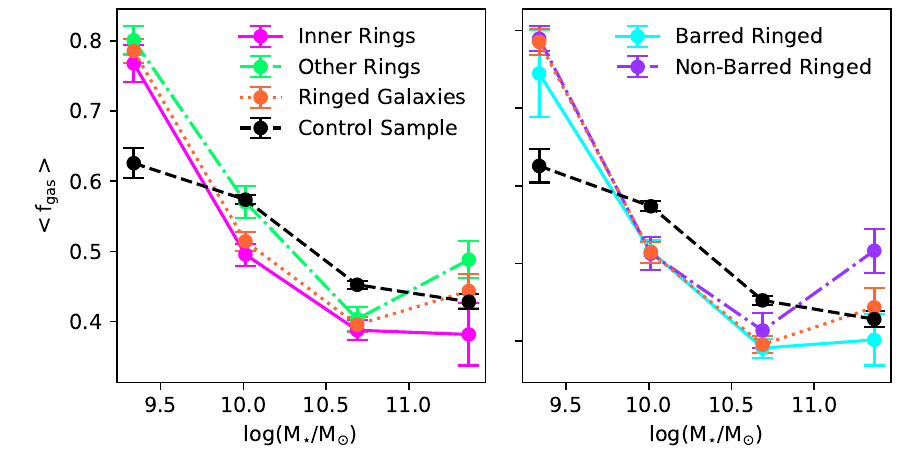}    
    \caption{Mean log(sSFR) with associated bootstrap errors for five bins in $f_{\rm gas}$ (upper panels) and mean values of $f_{\rm gas}$ with their associated bootstrap errors for five mass bins (lower panels). Left panels: Mean values for all ringed galaxies, inner ringed galaxies, other types of ringed galaxies, and control sample galaxies. Right panels: Mean values for barred ringed galaxies and non-barred ringed galaxies.}
    \label{fig:sfr-fgas}
\end{figure}

Examining the relationship between $f_{\rm gas}$ and the sSFR across different galaxy types can provide information into the role that certain structural characteristics play in galaxy evolution. Figure \ref{fig:sfr-fgas} shows how the mean values of $f_{\rm gas}$ and the log(sSFR) vary among galaxies with different rings and across various mass scales offering valuable information on the mechanisms regulating SF in galaxies. The upper panel of the figure shows that higher gas fractions generally correspond to increased sSFR, underscoring the critical role of gas availability in driving SF.
This figure also reveals that both inner ring galaxies and barred ring galaxies follow the same trend but to a lesser extent than their counterparts. Furthermore, in the lower panel of Fig. \ref{fig:sfr-fgas}, it is observed that galaxies with inner rings typically exhibit lower average $f_{\rm gas}$ in higher-mass ranges compared to those with other types of rings or the control sample. The same pattern is observed for barred ringed galaxies suggesting that galactic bars can affect the evolution of galaxies \citep{Kormedy2004} by redistributing their gas, potentially contributing to the cessation of SF \citep{Newnham}. The observed differences in $f_{\rm gas}$ between ringed and control galaxies are particularly noticeable at the lower end of the stellar mass range. However, the much smaller sample size for ringed galaxies, suggests that these results should be viewed with caution. The significantly higher gas fractions observed in ringed galaxies at lower masses may be a consequence of the limited number of objects in the bin rather than a true physical difference.

\subsection{Galaxy colors}

Utilizing the stellar magnitudes in the U, B, V, K, g, r, i, and z bands provided by TNG50, we investigated the colors of galaxies to better understand their evolutionary state and structural characteristics (\citealt{Feldmann,Pandey2020}). The magnitudes in the U, B, and V bands are based on Vega magnitudes and are described in \cite{Buser}. For the infrared K band, the IR detectors of the Palomar 200-inch telescope were used, also based on Vega magnitudes, while the magnitudes in the g, r, i, and z bands were calculated using the response functions of the SDSS camera, with an air mass of 1.3 measured in June 2001, and are expressed in AB magnitudes. Technical details and the implementation of these filters can be found in Section 3.2.1 of \cite{Stoughton}.

\begin{figure}[t]
    \centering
    \includegraphics[width=.49\textwidth]{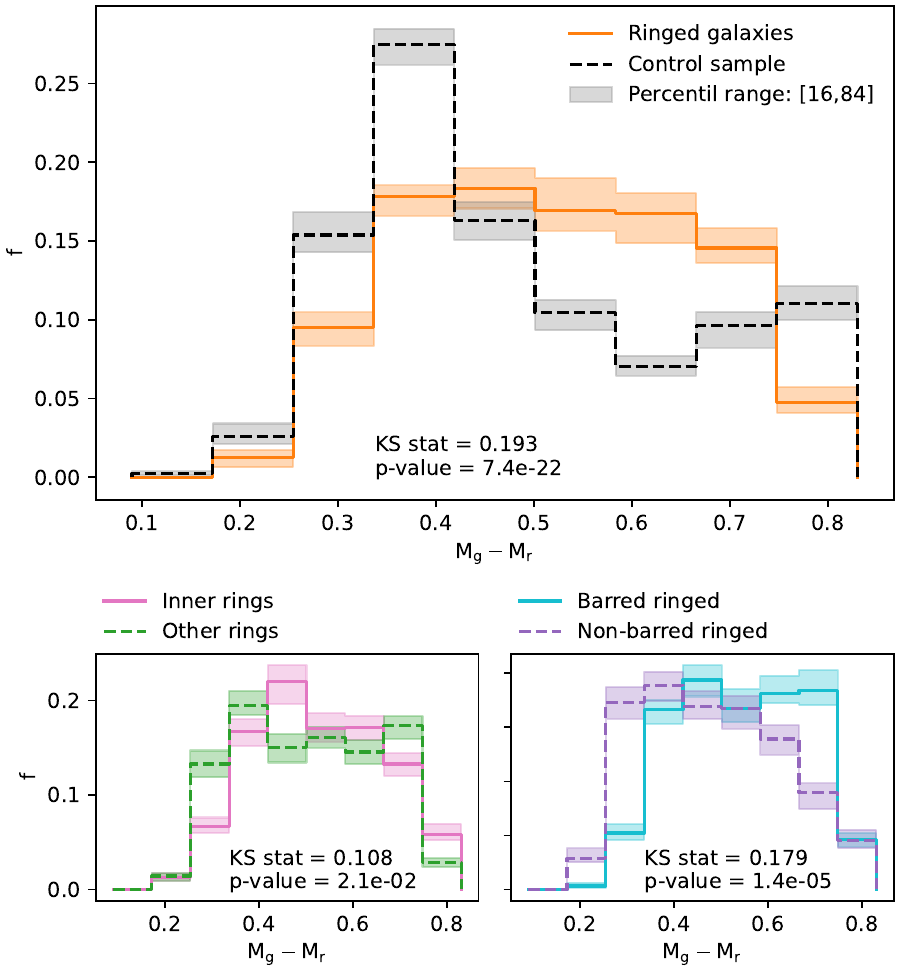}
    \caption{Color distribution $M_g - M_r$ for all analyzed $z$ in the TNG50 simulation. The upper panel shows the color fraction of ringed galaxies compared to the control sample. The lower panels offer a more detailed comparison, showing the color distribution of inner ringed galaxies versus other types of rings (left) and barred versus non-barred ringed galaxies (right) across the available $z$ in TNG50.}
    \label{fig:gr}
\end{figure}

We examined the distributions of $M_g-M_r$ between ringed galaxies and the control sample, as shown in Fig. \ref{fig:gr}. The color distribution of ringed galaxies shows a shift toward redder colors compared to the control sample. The peak of the distribution for ringed galaxies is notably lower in the $M_g-M_r$ index than the control sample. Suggesting a significant difference in stellar age and formation history compared to the control galaxies. In this context, ringed galaxies exhibit a more evolved and redder stellar population. This result is supported by the KS-test value, the lack of significant overlap between the bootstrap error bands of the distributions, and its consistency with the observational findings of \citetalias{Fernandez2021}, where galaxies with rings exhibited a tendency toward redder colors compared to galaxies without rings. This pattern was also found across different environments \citep{Fernandez2024}.
The lower-left panel of Fig. \ref{fig:gr} displays the distribution of color for galaxies with inner rings versus those with other types of rings. The data reveal that while all ringed galaxies generally tend to be redder, those with inner rings exhibit a slightly redder color index than galaxies with other types of rings. This could imply that galaxies with inner rings reduce gas more efficiently, leading to a faster transition to redder colors, especially those with greater mass, as observed in the previous section. Additionally, the comparison of colors between barred and non-barred ringed galaxies (lower-right panel) shows that barred ringed galaxies tend to have a redder color index than non-barred ones.

\begin{figure}[t]
    \centering
    \includegraphics[width=.45\textwidth]{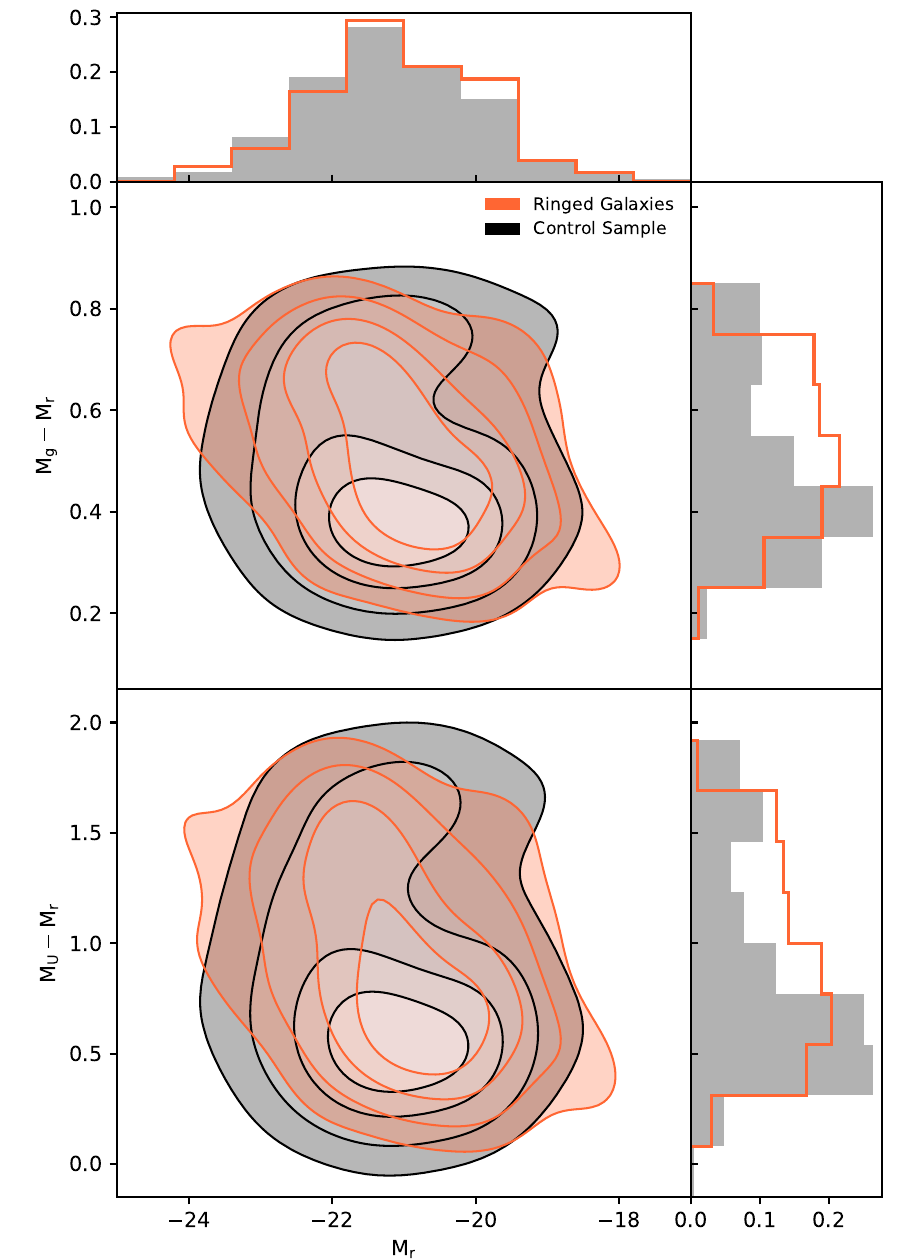}
    \caption{Color–magnitude diagrams. Upper panel: Contour plots of $M_g-M_r$ vs $M_r$ of ringed galaxies compared to the control sample. Lower panel: Contour plots of $M_U-M_r$ vs $M_r$ for the same samples. The contours enclose 25\%, 50\%, 75\%, and 90\% of the data for each sample.}
    \label{fig:color}
\end{figure}

In the top panel of Fig. \ref{fig:color}, the contour plot for $M_g-M_r$ versus $M_r$ is shown, where it can be seen that ringed galaxies are more concentrated in redder $M_g-M_r$ colors relative to the control sample. This indicates that ringed galaxies are generally more evolved, with denser and potentially older stellar populations clustering in redder colors at comparable magnitudes. This also suggests a maturity in the stellar populations or perhaps a greater efficiency in the aging processes within these galaxies.
Similarly, the contour plot presented in the bottom panel of Fig. \ref{fig:color} demonstrates that ringed galaxies tend to concentrate in redder $M_U-M_r$ indices, especially at brighter $M_r$ values. This trend again suggests that ringed galaxies likely host more mature stars or undergo processes that rapidly age their stellar constituents compared to the control sample.
These results show that ringed galaxies, might also represent stages where rapid stellar aging occurred, leading to a redder stellar population. This contrast in the color contours of ring galaxies compared to the control sample provides valuable information on how existing structural features, such as rings, might be part of processes that influence not only the current SF but also the aging of stellar populations within galaxies.

\subsection{Metallicity}

Gas metallicity serves as a crucial indicator of the chemical enrichment history of galaxies. In this work, metallicity was estimated using the oxygen-to-hydrogen abundance ratio, 12+log(O/H). Specifically, we utilized the atomic number abundance computed from the total mass fraction of each element in the gas cells which are star-forming, weighted by the cell SFR. This approach allows us to track the enrichment processes by SF and evolution within galaxies.

\begin{figure}[t]
    \centering
    \includegraphics[width=.49\textwidth]{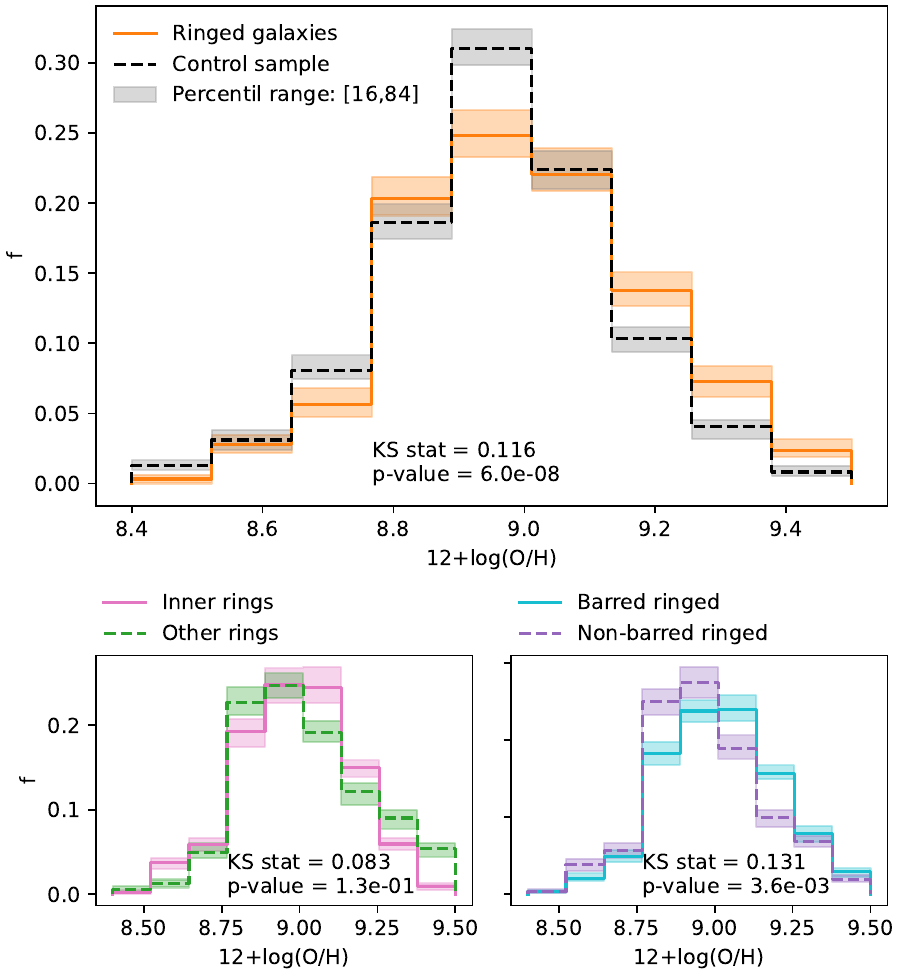}
    \caption{Distribution of 12+log(O/H) for all analyzed $z$ in the TNG50 simulation. The upper panel illustrates the metallicity fraction of ringed galaxies compared to the control sample. The lower panels show the metallicity distribution of inner ringed galaxies versus other types of rings (left) and barred versus non-barred ringed galaxies (right) across the available $z$ in TNG50.}
    \label{fig:metalicidad}
\end{figure}

Figure \ref{fig:metalicidad} displays the metallicity distribution for ringed galaxies compared to the control sample. The histogram reveals that ringed galaxies are slightly more metal-rich than those in the control sample. This observation is supported by the KS-test result, which shows a KS statistic of 0.116 and a p-value of $6\period0\times10^{-8}$. Furthermore, the range of dispersion of both distributions for the most metal-rich galaxies shows a significant separation, where at least 84\% of the ringed galaxies exhibit higher metallicities than 84\% of their non-ringed counterparts.

Expanding the analysis, Fig. \ref{fig:metalicidad} (bottom left panel) explores the differences in metallicity between galaxies with inner rings and those with other types of rings. The KS test results yield a statistic of 0.083 and a p-value of 0.130, indicating that while there is a difference in metallicity distribution between these two groups, it is not statistically significant.
The bottom right panel of the same figure displays the metallicity distribution between barred and non-barred ringed galaxies. The KS-test result shows a pronounced difference, with a KS-statistic of 0.131 and a p-value of 0.004, suggesting that barred ringed galaxies have a significantly different and higher metallicity distribution compared to non-barred ringed galaxies. This could reflect the role of the bar in driving gas toward the central regions, thereby enhancing initial central SF and subsequent metal enrichment. Furthermore, this result is consistent with the findings of \citetalias{Fernandez2021} for barred ringed galaxies.

\begin{figure}[t]
    \centering
    \includegraphics[width=.49\textwidth]{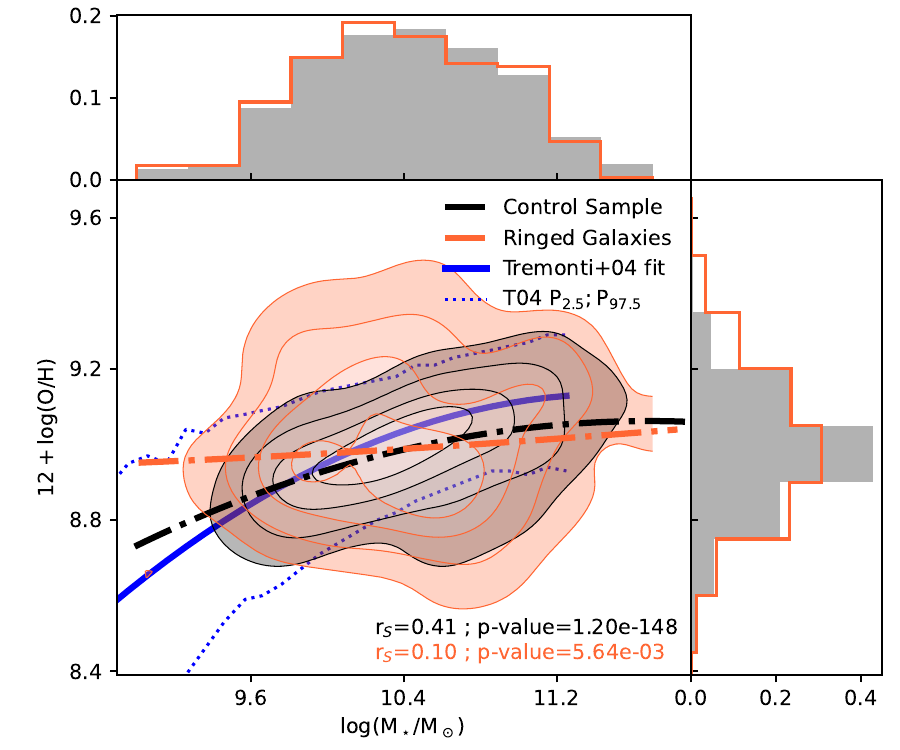} 
    \caption{Mass-metallicity diagram and distributions of mass and metallicity for ringed galaxies (orange) and control sample galaxies (black). The contours enclose 25\%, 50\%, 75\%, and 90\% of the data for each sample. The dash-dotted lines are degree 2 polynomial fits calculated using a Theil-Sen regression, while the solid blue line shows the relationship obtained by \cite{Tremonti2004} and the dotted lines enclose 95\% of the SDSS data used in said work.}
    \label{fig:masmet}
\end{figure}

Finally, Fig. \ref{fig:masmet} shows the mass-metallicity relation (MZR) for ringed galaxies and the control sample. The figure reveals that ringed galaxies exhibit greater variability in metallicity compared to non-ringed galaxies at a same $M_\star$.
Most of the control sample members $(\sim 90\%)$ are in good agreement with the relation found by \cite{Tremonti2004}, denoted by the solid blue line, who used SDSS imaging and spectroscopy to study the relationship between stellar mass and gas-phase metallicity of $\sim 53000$ star-forming galaxies at $z \sim 0\period1$, where 95\% of these data are contained by the dotted blue lines.
However, this does not occur for ringed galaxies, which exhibit a weak correlation between metallicity and $M_\star$, as shown by the calculated Spearman coefficient of $r_S = 0 \period 10$ (p-value$=5\period64 \times 10^{-3}$).
To compare the trends of both samples, we estimated polynomial fits of degree 2 using a Theil-Sen regression\footnote{The Theil-Sen estimator uses a generalization of the median in multiple dimensions, making it robust to multivariate outliers in low-dimensional problems.} (dash-dotted lines). We observe a relevant fraction of high-mass galaxies with low metallicity in TNG50 that produce a smoother MZR in contrast to the observed one.
However, at low mass, the ringed galaxies exhibit a considerable bias to higher metallicities, resulting in strongly flattened mass-metallicity relationships.

In analyzing galaxies from the Mapping Nearby Galaxies at Apache Point Observatory (MaNGA) survey \citep{Bundy2015} and the Evolution and Assembly of GaLaxies and their Environments (EAGLE) simulation \citep{Schaye2015}, \cite{Jara2024} found an additional dependence of the MZR on metallicity gradients $(\nabla(O/H))$, with two different regimes emerging for galaxies with masses lower or higher than log$(M_\star/M_\odot) \sim 9\period75$. Low-mass galaxies with a strong negative $\nabla(O/H)$ are consistently more enriched than the median MZR and show high SF activity, regardless of surface star density. In addition, \cite{Tissera2016} observed that large dispersion in metallicity gradients as a function of $M_\star$ could be ascribed to the effects of dynamical processes as well as those regulating the conversion of gas into stars. In this context, our results might suggest the presence of distinct evolutionary processes affecting ringed and non-ringed galaxies.

\section{Rings features}

\subsection{Radial profiles}

\begin{figure*}[t]
    \center
    \includegraphics[trim={0 25 5 0},width=.99\textwidth]{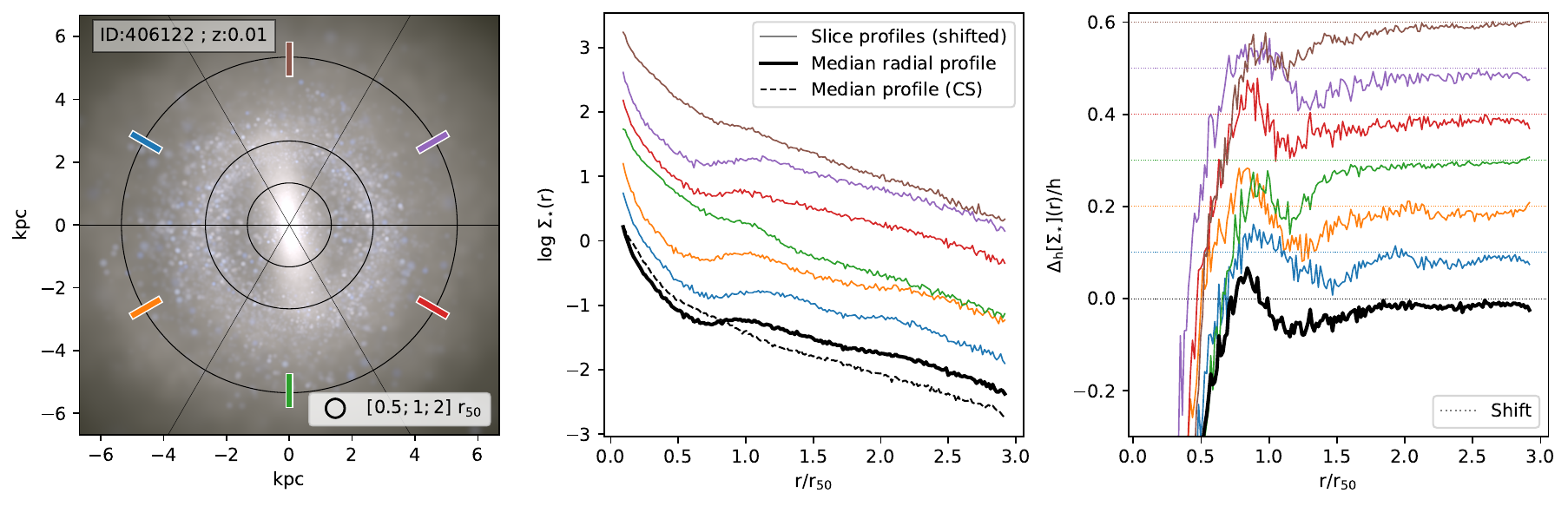}
    \includegraphics[trim={5 25 5 0},width=.99\textwidth]{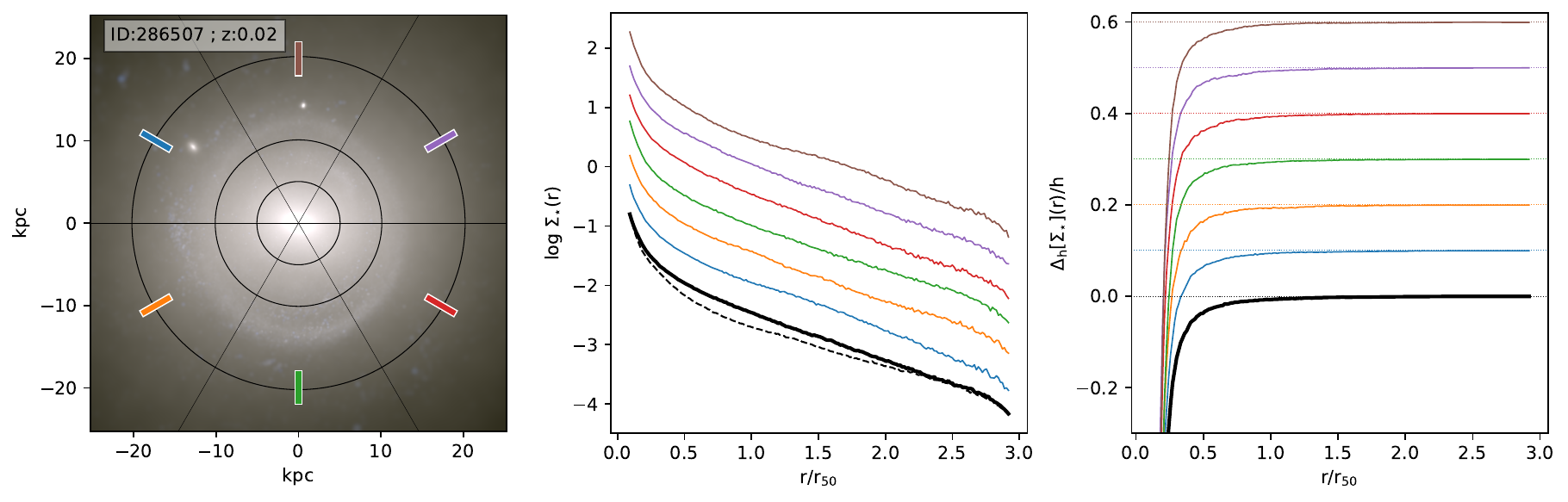}
    \includegraphics[trim={8 12 5 0},width=.99\textwidth]{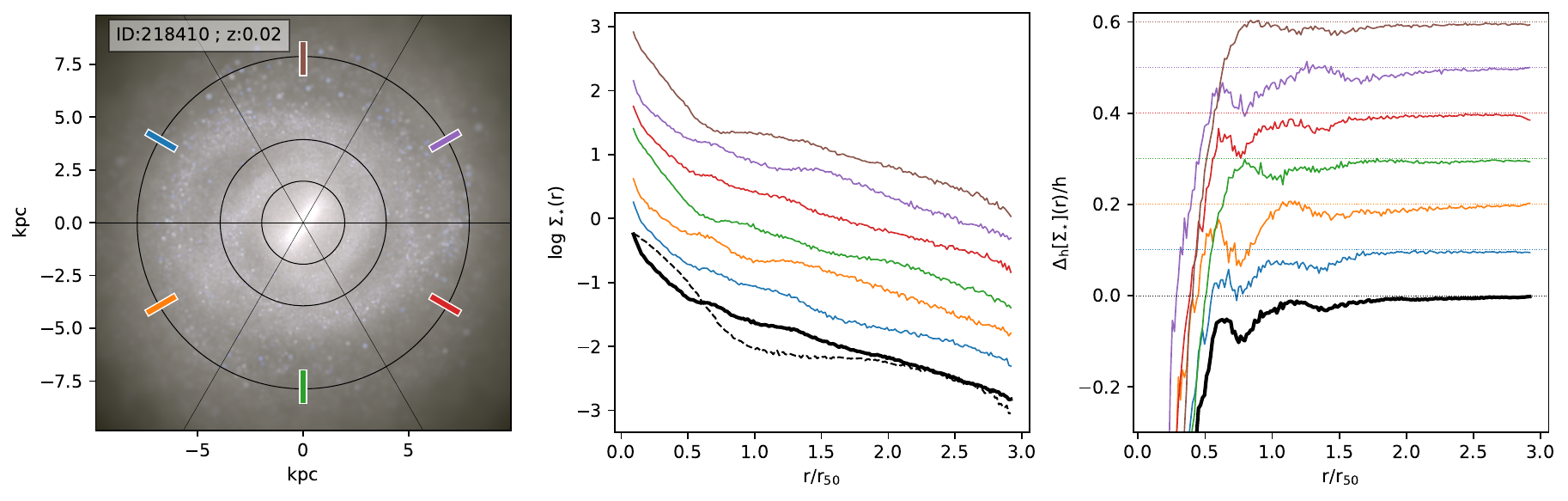}
    \caption{Examples of radial surface mass density profiles of ringed galaxies. The first column shows the face-on projection of a galaxy with an inner ring (top row), an outer ring (middle row), and i+o rings (bottom row). The polar mesh that defines the azimuthal sections where the radial profiles are calculated is also displayed. These profiles are presented in the middle column, identified with the color that denotes their sections in the grid, and are shifted along the y-axis. The global profiles of the ringed galaxy and its control group are represented by solid and dashed black lines. The right column displays the derivatives of the profiles, approximated by forward finite differences.}
    \label{fig:ring_profiles}
\end{figure*}

Following the visual identification and classification of the ringed galaxies in TNG50 and the study of their global properties in comparison with similar systems without rings, the samples obtained allow us to quantitatively characterize the structural features of the different classes of rings. With this aim, we compute the surface density of stellar mass projected in the galactic plane as a function of the distance to the barycenter, $\Sigma_\star(r)$. \cite{WandLily} show that TNG50 galaxies belonging to the star formation main sequence (SFMS; with $\rm sSFR\gtrsim 10^{-12}\ yr^{-1}$ and  log$(M_\star/M_\odot h^{-1}) > 9\period5$), exhibit radial profiles that well describe exponential disks with a Sérsic core (or bulge) in the galactic center. This seems to be in good agreement with observations on the stellar distribution of disk galaxies \citep{PohlenTrujillo2006, Meert2013}.

The presence of a ring in the disk of the galaxy can be identified by a bump in its exponential profile, which can also present distinctive characteristics corresponding to the three different general types defined by \cite{PohlenTrujillo2006} and \cite{Erwin2008}, depending on the observed break. To prevent compact and massive structures present in the galactic disk from originating a signal in a global radial profile that could be confused with the existence of a ring, or hiding said signal, we divided the front projection of the galaxy into six identical triangular sections, adjacent and non-overlapping, and we calculated the surface mass density profile in each section.

Examples of the result of this procedure can be seen in Fig. \ref{fig:ring_profiles}, for a galaxy with an inner ring (upper row), with an outer ring (middle row) and with i+o rings (lower row). In the first column we can see the front perspective of each type of galaxy, on which the grid used to compute the mass profiles is shown. These are presented in the middle column, in units of the $r_{50}$ of the galaxy, and are identified with the colors that denote the different sections of the grid. To facilitate clear observation, each profile is slightly displaced from the rest on the ordinate axis. The global profiles of the ringed system and its control group galaxies are represented by solid and dashed black lines.

Particularly for barred galaxies, we observe that profiles mapping such structures present a smoother decay compared to those that are not traversed by the bar. By constructing a general profile by means of the median, along the radius, of the different azimuthal contributions calculated for each galaxy, the predominance of any of them due to their mass concentration can be blurred.

In these examples, especially in galaxies with an inner ring that is more robust, we can notice how the profiles exhibit a prominence in the radius where these structures are visualized. Such alteration can be identified as a marked change of the local slope of the profile in the radial environment close to the ring. Then, we can approximate this derivative by using forward finite differences\footnote{$\rm \Delta_{h}[f](x) = f(x+h) - f(x)$}. The resulting curves $\rm \Delta_{h}[\Sigma_\star](r)/h$\footnote{$\rm \Delta_{h}[\Sigma_\star]/h = \left[ \Sigma_\star(r+h) - \Sigma_\star(r) \right]/h$}, are shown in the right column of Fig.~\ref{fig:ring_profiles} (also slightly displaced in the value shown by the horizontal dashed lines), and in them it is seen how the conglomerate of stars that form the ring produces flatter or positive slopes in the radial profiles.

\subsection{Formats of the rings}

\begin{figure*}[t]
    \center
    \includegraphics[trim={0 7 0 0}, width=1\textwidth]{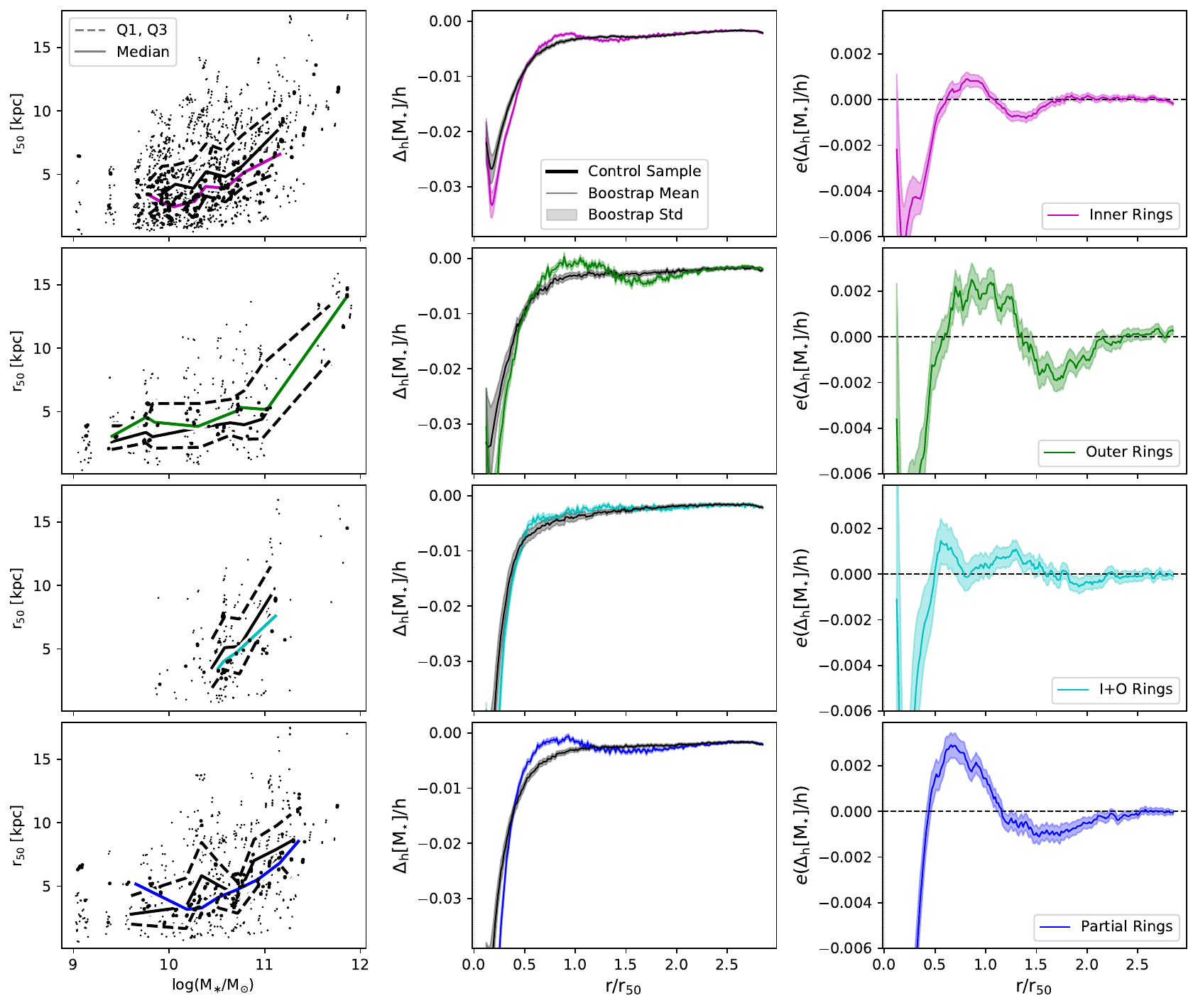}
    \caption{Structural description of the rings. Left column: mass-size relation for the different types of ringed galaxies (inner, outer, i+o, and partial rings) with their respective control samples; The solid lines show the median trends for each sample, constructed from a stellar masses binning, while dashed lines enclosed the inter-quartile range of each control distribution. Middle and right columns: Bootstrap mean and standard deviation of the stack, as a function of $r_{50}$, of the slopes of the radial mass profile and residuals obtained by approximating the slope profile of a ringed galaxy by the median of those of its control group.}
    \label{fig:slope_radial_profiles}
\end{figure*}

\cite{WandLily} used the $r_{50}$ as a proxy for galactic size and found that, for TNG50 galaxies at $z \sim 0$, the general trend of the mass-size relation is in good agreement with that of observations over four orders of magnitude in stellar mass, although with an excess of about 0.11 dex at the high-mass end, $M_\star > 10^{10} M_\odot\ \rm h^{-1}$. The left column of Fig.~\ref{fig:slope_radial_profiles} displays the mass-size relation for the different types of ringed galaxies with their respective control samples. We observe that, in general, the median $r_{50}$ of outer ring galaxies is similar to or slightly larger than that of their control group at the same mass. However, for galaxies with an inner or partial ring the opposite occurs, while for systems with i+o rings the comparison is diffuse due to the low number of these galaxies in our sample.

Therefore, expressing $\Sigma_\star(r)$ in terms of $r_{50}$ allows us to stack the profiles and their approximate derivatives for galaxies of different masses. This approach roughly aligns the different radial regions relative to the size of each galaxy, enhancing the signal from the annular structures under study. As a consequence, it defines a characteristic undulation on the bootstrap mean of the radial rate of change of the mass profiles of ringed galaxies $\Delta_{\rm h}[M_\star](\frac{r}{r_{50}})/\rm h$\footnote{$\Delta_{\rm h}[M_\star](\frac{r}{r_{50}})/ \rm{h} = \left[ M_\star(\frac{r}{r_{50}}+h)-M_\star(\frac{r}{r_{50}}) \right]/\rm h$} that is not present in the control sample, as displayed in the second column of Fig.~\ref{fig:slope_radial_profiles}. In order to highlight that oscillation, the right column shows the bootstrap mean and spread of the residuals, 
\begin{equation}
e = \left[ \Delta_{\rm h}[M_{\star,\rm{RG}}](\frac{r}{r_{50}}) - \rm{Me} \left(\Delta_{\rm h}[M_{\star,\rm{CS}}](\frac{r}{r_{50}}) \right) \right]/ \rm h
\end{equation}
obtained by approximating the $\Delta_{\rm h}[M_\star](\frac{r}{r_{50}})/\rm h$ of a ringed galaxy by the median slope profile of its control group.

The central deficit exhibited by the residues is due to the region of low-mass density present before the ring structure. This region generates a more abrupt decline in the slope of the radial profile, which then tends to stabilize and rapidly increases as it approaches the ring.
Here we observe that the inner rings are located approximately at $r_{50}$ while the outer rings at 1.5 $r_{50}$. The small variation between their relative positions reflects the relevant contribution of the ring mass to establishing the $r_{50}$. Both contributions tend to be compensated in galaxies with i+o rings, from which we can infer that the inner rings are more compact and massive than the outer rings because the wave generated by the former has a shorter length and is closer to $r_{50}$. Finally, the curves obtained for galaxies with partial rings show deeper mass profiles than those of their control group, in the central region of the galaxy, where the arc structures or pseudo-rings are located beyond $r_{50}$, extending as much or more than the external rings.

As a first comparison with observations, we resort to the catalog of rings detected in the Spitzer Survey of Stellar Structure in Galaxies (S$^4$G), generated by \citet[][Arrakis]{Comeron2014}. This catalog contains data on 724 ringed galaxies including ring sizes relative to galaxy sizes, where the latter are represented by the diameter at the level $\rm \mu_{3 \period 6\mu m} = 25 \period 5\ mag\ arcsec^{-2}$ ($D_{25} = 2 \times R_{25}$). \citet{Leroy2021}, using the Physics at High Angular resolution in Nearby GalaxieS (PHANGS-ALMA) survey, proposed the relation $r_{50} \approx 3 \period 08 R_{25}$ and found that it produces a median $r_{50}$ approximately 4\% larger than that obtained by \citet[][S$^4$G data release]{Muñoz-Mateos2015}, with a dispersion of 25\% $(1\sigma)$. Applying this conversion we obtained that the inner rings are located at $0 \period 84 \pm 0 \period 37 \ r_{50}$ while the outer rings are at $1 \period 84 \pm 0 \period 58 \ r_{50}$. These locations agree well with our results in TNG50.

\section{Summary and conclusions}

We present a catalog of ringed galaxies selected from the TNG50 simulation. We focus on galaxies within a redshift range ($0 \period 01 < z < 0 \period 1$) comparable to the SDSS-DR14 ringed galaxy catalog presented by \citetalias{Fernandez2021}. TNG50 galaxies were also selected with $M_\star > 10^9 M_\odot$, $\rm{log(sSFR/yr}^{-1}) > -13$, and $r_{50} > 1$ kpc following \citetalias{Fernandez2021}. We conducted a visual classification through the analysis of synthetic, dust-free images of face-on galaxies, where the primary parameter considered was the presence of ring structures. These structures were classified into inner rings, outer rings, i+o rings, and pseudo-rings. We find that 807 galaxies within the simulation met these criteria and were classified as ringed galaxies. Approximately 59\% of these galaxies possess an inner ring, 22\% a partial ring, 12\% an outer ring, and 7\% i+o rings.

The analysis of the distribution of ring types across different redshifts reveals that at low values ($z \thicksim 0 \period 01$), galaxies with inner rings are predominant (74\%), while this fraction slightly decreases as $z$ increases to $z \thicksim 0 \period 1$ (54\%). Outer rings and i+o rings  maintain a stable distribution across $z$, indicating a less time-dependent evolution. In contrast, galaxies with partial rings increase in frequency from 3\% at $z \thicksim 0 \period 01$ to 25\% at $z \thicksim 0 \period 1$, suggesting that these rings are temporary features sensitive to galactic dynamical processes. This trend also suggests that partial rings are more common in the early evolutionary stages of galaxies and may transform into more stable outer rings over time, reflecting a process of structural maturation in galaxies.
We also identified that 64\% of our ringed galaxies have bars. We observed an increase in the number and percentage of galaxies with rings and bars as $z$ increases, indicating that bars may be more consolidated at early stages of galactic development \citep{RosasGuevara}. In contrast, ringed galaxies without bars do not show such a pronounced increase.

We compared our percentage results with the observational sample generated by \citetalias{Fernandez2021}  
and find that the simulated galaxies show a higher prevalence of inner rings (59\% in TNG50 vs. 49\% in SDSS). However, i+o rings are less common in the simulation (7\% in TNG50 vs. 21\% in SDSS).
Partial rings exhibit a slight variation, appearing more frequently in TNG50 at 22\% compared to 19\% in SDSS. In contrast, outer rings display similar percentages in both catalogs.
Additionally, the proportion of galaxies with bars is slightly lower in TNG50 (64\%) compared to SDSS (68\%). These differences may reflect variations in the underlying galactic dynamics or in the resolution of simulations and observations.

We also constructed a control sample of non-ringed galaxies with similar $z$, $M_\star$, and environmental density distributions to those of ringed galaxies, in order to study the properties of ringed galaxies in the simulation. Additionally, we divided our ringed galaxies into four groups for this analysis: (i) ringed galaxies with an inner ring; (ii) ringed galaxies with other types of rings, including double rings (i+o), outer rings, and partial rings; (iii) barred ringed galaxies; and (iv) unbarred ringed galaxies. From the study of these properties, we find that:
\begin{itemize}
    \item Ringed galaxies tend to have a lower SFR compared to the control sample, with those featuring inner rings and bars exhibiting the lowest rates. Additionally, barred ringed galaxies consistently show a lower sSFR across most mass ranges, underscoring the influence of morphological features on SF dynamics. The $f_{\rm gas}$ provides further insight, with lower fractions indicating more evolved galaxies. Analyzing $f_{\rm gas}$ and log(sSFR) across different galaxy types reveals that higher gas fractions correspond to an increased SFR, yet inner rings and bars still lead to lower rates, suggesting that galaxies with these types of structures may exert some influence on the regulation of SF and the overall evolution. 
    \item The analysis of galaxy color properties reveals insights into SF, galactic evolution, and environmental influences. Using TNG50 stellar magnitudes across various bands, we investigated galaxy colors to understand their evolutionary state and structural characteristics. We find that ringed galaxies display a shift toward redder colors compared to the control sample, indicating older, more evolved stellar populations. Specifically, galaxies with inner rings are slightly redder than those with other types of rings, suggesting a more efficient gas reduction, particularly in the higher-mass ones. Barred ringed galaxies also exhibit redder color indices than non-barred ones. Contour plots show that ringed galaxies cluster in redder $M_g - M_r$ and $M_U - M_r$ indices, especially at brighter magnitudes, highlighting their maturity and the rapid aging processes of their stellar populations. 
    \item The analysis of the metallicity reveals that ringed galaxies are slightly more metallic compared to the non-ringed ones, suggesting that these galaxies have undergone diverse evolutionary stages and enrichment processes. Furthermore, it was found that barred ringed galaxies have a higher metallicity compared to non-barred ones. This finding may reflect the role of the bar in driving gas toward the central regions, thereby enhancing initial SF and subsequent metal enrichment.
    Additionally, regarding the MZR, ringed galaxies exhibit greater variability in metallicity for a given $M_\star$ compared to the control sample. Ringed galaxies show a weak correlation between metallicity and $M_\star$, with a tendency toward higher metallicity at lower masses, suggesting distinct evolutionary processes compared to non-ringed galaxies.
\end{itemize}

It is important to add that many of these results are consistent with the results obtained observationally by \citetalias{Fernandez2021}. Finally, we studied the characteristics of ringed galaxies through radial profiles of projected stellar mass surface density, $\Sigma_\star(r)$. We find that the resulting profiles reveal prominences at the radii where the rings are located, particularly in galaxies with robust inner rings, and show how these stellar conglomerates affect the local slope of the profile.

From the study of galactic size, measured by $r_{50}$, we find that galaxies with outer rings have $r_{50}$ values similar to or slightly greater than their control group, while those with inner or partial rings have smaller sizes. Expressing $\Sigma_\star(r)$ in terms of $r_{50}$ allows for the alignment of radial regions and enhances the signal of the ring structures. The residuals in the profiles show a central deficit before the ring, with inner and outer rings located approximately at $r_{50}$ and 1.5 $r_{50}$, respectively. Galaxies with both i+o rings indicate that the inner rings are more compact and massive. Galaxies with partial rings present deeper mass profiles in the central region compared to their control group, with pseudo-rings extending beyond $r_{50}$.

Our comprehensive analysis of the TNG50 simulation has provided valuable insights into the structure and evolution of ringed galaxies. By examining their properties, stellar mass surface density profiles, and comparing their sizes, we have highlighted the role of ringed structures in reflecting the underlying processes that influence galactic dynamics. Our findings align well with observational data, reinforcing the idea that the interplay between structural features, such as rings and bars, and the processes that drive galaxy formation and evolution is complex. Understanding these relationships is crucial to advancing our knowledge of galactic dynamics and the factors that regulate their properties over time.

\begin{acknowledgements}

This work was partially supported by the Consejo Nacional de Investigaciones Cient\'{\i}ficas y T\'ecnicas and the Secretar\'{\i}a de Ciencia y T\'ecnica de la Universidad Nacional de San Juan.
ES acknowledges funding by Fondecyt-ANID Postdoctoral 2024 Project N°3240644 and thanks the N\'ucleo Milenio ERIS.
PBT acknowledges partial funding by Fondecyt Regular 2024 Project N°1240465, ANID BASAL project FB210003 and also thanks N\'ucleo Milenio ERIS. This project has received funding from the European Union Horizon 2020 Research and Innovation Programme under the Marie Sklodowska-Curie grant agreement No 734374 - LACEGAL. We acknowledge the use of the Ladgerda Cluster (Fondecyt 1200703/2020 hosted at the Institute for Astrophysics, Chile)
\end{acknowledgements}

\bibliographystyle{aa} 
\bibliography{biblio} 

\end{document}